\documentclass{jfm}
\pdfoutput=1
\usepackage{amsmath}
\usepackage{amssymb}
\usepackage{graphicx}
\usepackage{color}
\usepackage[applemac]{inputenc}
\usepackage{array}
\usepackage[left]{lineno}
\usepackage[usenames,dvipsnames]{xcolor}
\usepackage{tikz}
\usetikzlibrary{arrows}
\usepackage{natbib}

\ifCUPmtlplainloaded \else
  \checkfont{eurm10}
  \iffontfound
    \IfFileExists{upmath.sty}
      {\typeout{^^JFound AMS Euler Roman fonts on the system,
                   using the 'upmath' package.^^J}%
       \usepackage{upmath}}
      {\typeout{^^JFound AMS Euler Roman fonts on the system, but you
                   dont seem to have the}%
       \typeout{'upmath' package installed. JFM.cls can take advantage
                 of these fonts,^^Jif you use 'upmath' package.^^J}%
      }
  \else
  \fi
\fi

\ifCUPmtlplainloaded \else
  \checkfont{msam10}
  \iffontfound
    \IfFileExists{amssymb.sty}
      {\typeout{^^JFound AMS Symbol fonts on the system, using the
                'amssymb' package.^^J}%
       \usepackage{amssymb}%
         \let\leq=\leqslant
         \let\geq=\geqslant
      }{}
  \fi
\fi

\ifCUPmtlplainloaded \else
  \IfFileExists{amsbsy.sty}
    {\typeout{^^JFound the 'amsbsy' package on the system, using it.^^J}%
     \usepackage{amsbsy}}
    {}
\fi

\newsavebox{\astrutbox}
\sbox{\astrutbox}{\rule[-5pt]{0pt}{20pt}}

\title[]{Experiments on the fragmentation of a buoyant liquid volume in another liquid}

\author[M. Landeau, R. Deguen and P. Olson]%
{M. Landeau$^{1,2}$
\thanks{Email address for correspondence: mlandeau@jhu.edu},\ns
R. Deguen$^{3,4}$\break and P. Olson$^1$ 
}

\affiliation{
$^1$Department of Earth and Planetary Sciences, Johns Hopkins University, Baltimore, MD 21218, USA\\[\affilskip]
$^2$Dynamique des Fluides Géologiques, Institut de Physique du Globe de Paris,
Sorbonne Paris Cité, Université Paris Diderot, CNRS UMR 7154, 1 rue Jussieu 75238 Paris cedex 5, France 
\\[\affilskip]
$^3$Laboratoire de Géologie de Lyon, Université Lyon 1, 2 rue Rapha\"el Dubois, 69622 Villeurbanne, France
\\[\affilskip]
$^4$Institut de Mécanique des Fluides de Toulouse, Université de Toulouse (INPT, UPS) and CNRS, 2 allée Camille Soula, Toulouse, 31400, France.\\[\affilskip]
}

\pubyear{2010}
\volume{650}
\pagerange{119--126}
\date{}

\newcommand\Rey{\mbox{\textit{Re}}}
\newcommand\Web{\mbox{\textit{We}}}
\newcommand\Bo{\mbox{\textit{Bo}}}
\newcommand\DP{\mbox{\textit{P}}}
\newcommand\Oh{\mbox{\textit{Oh}}}

\newcommand{\coloursquare}[1]{\tikz \path [fill=#1,draw=#1,line width = 1pt] (0,0) rectangle (6pt,6pt) ;}
\newcommand{\greysquare}[1]{\tikz \path [fill=gray,draw=#1,line width = 1pt] (0,0) rectangle (6pt,6pt) ;}
\newcommand{\bicoloursquare}[2]{\tikz \path [fill=#1,draw=#2,line width = 0.6pt] (0,0) rectangle (6pt,6pt) ;}
\newcommand{\emptysquare}[1]{\tikz \path [fill=white,draw=#1,line width = 1pt] (0,0) rectangle (6pt,6pt) ;}
\newcommand{\colourcircle}[1]{\tikz \path [fill=#1,draw=#1,line width = 1pt] (0,0) circle (3pt) ;}
\newcommand{\bicolourcircle}[2]{\tikz \path [fill=#1,draw=#2,line width = 0.6pt] (0,0) circle (3pt) ;}

\newcommand{\greycircle}[1]{\tikz \path [fill=gray,draw=#1,line width = 1pt] (0,0) circle (3pt) ;}
\newcommand{\colourtriangle}[1]{\tikz [line width = 5pt] \draw[color=#1] (0,0.03) --  (0.03,0.03) -- (0.015,0) -- cycle;}
\newcommand{\bicolourtriangle}[2]{\tikz [line width = 0.6pt] \draw[color=#2, fill =#1] (0,0.2) --  (0.2,0.2) -- (0.1,0) -- cycle;}
\newcommand{\bicolourtriangleup}[2]{\tikz [line width = 0.6pt] \draw[color=#2, fill =#1] (0,0) --  (0.1,0.2) -- (0.2,0) -- cycle;}
\newcommand{\colourtriangleup}[1]{\tikz [line width = 5pt] \draw[color=#1] (0,0) --  (0.015,0.03) -- (0.03,0) -- cycle;}
\newcommand{\greytriangleup}[1]{\tikz [line width = 1pt] \draw[color=#1, fill =gray] (0,0) --  (0.1,0.2) -- (0.2,0) -- cycle;}
\newcommand{\emptytriangle}[1]{\tikz [line width = 1pt] \draw[color=#1] (0,0.2) --  (0.2,0.2) -- (0.1,0) -- cycle;}
\newcommand{\greytriangle}[1]{\tikz [line width = 1pt] \draw[color=#1, fill = gray] (0,0.2) --  (0.2,0.2) -- (0.1,0) -- cycle;}
\newcommand{\emptydiamond}[1]{\tikz [line width = 1pt] \draw[color=#1] (0,0.12) --  (0.1,0.24) -- (0.2,0.12) -- (0.1,0) -- cycle;}
\newcommand{\greydiamond}[1]{\tikz [line width = 1pt] \draw[color=#1, fill = gray] (0,0.12) --  (0.1,0.24) -- (0.2,0.12) -- (0.1,0) -- cycle;}

\newcommand{\bicolourtrianglec}[2]{\tikz{\path [fill=#2,draw=#1,line width = 0.6pt] (0,0) circle (4pt);
 \draw[color=#1, fill =#2] (-0.1,0.06) --  (0.1,0.06) -- (-0.,-0.1) -- cycle;} 
}
\newcommand{\bicoloursquarec}[2]{\tikz{\path [fill=#2,draw=#1,line width = 0.6pt] (0,0) circle (4pt);
 \draw[color=#1, fill =#2] (-0.07,-0.07) --  (0.07,-0.07) -- (0.07,0.07) -- (-0.07,0.07) -- cycle;} 
}

\newcommand\Star[3][]{\tikz %
\path[#1] (0  :#3) -- ( 36:#2) 
       -- (72 :#3) -- (108:#2)
       -- (144:#3) -- (180:#2)
       -- (216:#3) -- (252:#2)
       -- (288:#3) -- (324:#2)--cycle;}

\newlength{\mahaut}
\newlength{\malarg}

\newcommand{\intevariable}[3]{
\settowidth{\malarg}{$\displaystyle \int $}
\settodepth{\mahaut}{$\displaystyle #3 $}
\setlength{\mahaut}{1.2\mahaut}
{\resizebox{\malarg}{\mahaut}{$\displaystyle\int$}}_{\kern-0.5\malarg#1}^{#2}
#3
}

\begin{document}
\maketitle

\begin{abstract}

%
%
%
%
%

We present experiments on the instability and fragmentation of volumes of heavier liquid released into lighter immiscible liquids. We focus on the regime defined by small Ohnesorge numbers, density ratios of order one, and variable Weber numbers. The observed stages in the fragmentation process include deformation of the released fluid by either Rayleigh-Taylor instability or vortex ring roll-up and destabilization, formation of filamentary structures, capillary instability, and drop formation. At low and intermediate Weber numbers, a wide variety of fragmentation regimes is identified. Those regimes depend on early deformations, which mainly result from a competition between the growth of Rayleigh-Taylor instabilities and the roll-up of a vortex ring. At high Weber numbers, turbulent vortex ring formation is observed. We have adapted the standard theory of turbulent entrainment to buoyant vortex rings with initial momentum. We find consistency between this theory and our experiments, indicating that the concept of turbulent entrainment is valid for non-dispersed immiscible fluids at large Weber and Reynolds numbers. 

\end{abstract}


\section{Introduction}

Buoyancy-driven fragmentation of one liquid in another immiscible liquid 
likely occurred on a massive scale during the formation of the terrestrial planets. For example, it is thought that Earth acquired much of its present mass through high speed collisions between planetary embryos, in which both of the impacting objects consisted of a silicate mantle and a metallic core \citep{Melosh1990,Yoshino2003,Schersten2006,Ricard2009}. The enormous energy release following these impacts resulted in prodigious melting, creating an environment in which dense liquid metal blobs fell and subsequently fragmented within deep molten silicate magma oceans \citep{TonksMelosh1993,Pierazzo1997}. Less violent but still dramatic present-day analogs of this phenomenon include releases of petroleum into the ocean through well discharges, such as occurred in 2010 during the Deepwater Horizon disaster \citep{McNutt2012,Reddy2012,Camilli2012}.

Most fluid fragmentation processes involve a regular sequence of steps, including deformation 
or destabilization of the initial mass, formation of filamentary structures called liquid ligaments, breakup of ligaments usually involving capillary instabilities \citep[e.g.][]{Hinze1955,MarmottantVillermaux2004,VillermauxBossa2009}. The destabilizing mechanisms generally set the mean size of the resulting drops, whereas ligament dynamics plays a dominant role in determining the resulting drop size distribution \citep{MarmottantVillermaux2004,BremondVillermaux2006,VillermauxBossa2009,VillermauxBossa2011}.    
A principal control parameter in any fluid fragmentation process is the Weber number $We$, which measures the relative importance of the dynamic pressure and the capillary restoring pressure. 
Breakup, the final fragmentation stage, is usually divided in \textit{primary} and \textit{secondary breakup}. Primary breakup refers to the stage where the initial liquid volume 
divides into several disconnected blobs or drops. If the Weber number of the resulting blobs (based on the blob size and blob velocity) is larger than the critical value for breakup $We_c$, secondary breakups occur. The critical Weber number $We_c$ is generally of order $10$ 
but it varies with the flow regime in the surrounding fluid, especially with the Reynolds number \citep{Hinze1955}. 
Another main control parameter is the Ohnesorge number $Oh$, which measures the importance of viscous forces versus interfacial forces and inertia. 
Very large Weber and Reynolds numbers and $Oh$ much smaller than $1$ are the relevant regimes for planetary formation.
 
Fragmentation of a finite volume of liquid at low $Oh$ has been extensively studied in air 
\citep[reviewed in][]{PilchandErdman,Faeth1995,Gelfand1996,Guildenbecher2009,Theofanous2011}. A rich variety of fragmentation regimes has been identified, including \textit{vibrational breakup}, \textit{bag breakup}, \textit{multimode breakup}, \textit{shear breakup}, \textit{catastrophic breakup} (the terminology varies from one study to the other). Recently, \cite{Theofanous2004} and \cite{Theofanous2008} have proposed another categorization based on only two main fragmentation regimes : 
the \textit{Rayleigh-Taylor (RT) piercing} regime, 
in which early deformations result from Rayleigh-Taylor instabilities (RTI), 
which appear when an interface between two fluids of different density is subjected to an acceleration directed towards the lighter fluid, 
and the \textit{shear-induced entrainment} regime, 
interpreted as the suppression of RTI due to straining motions associated with the global shear. 
In general, the Weber number is the main control parameter governing transitions between the different fragmentation regimes.

%

Fragmentation of a buoyant liquid volume at density ratio of order one (i.e. in a liquid-liquid system) has received less attention. The maximum Weber numbers reached in three dimensional numerical simulations \citep{Ichikawa2010} of the breakup of drops falling in another immiscible liquid is about $10-15$. 
Axisymmetric simulations reach higher Weber numbers and are useful to compute the early deformations of a blob falling under gravity \citep{HanTrygg1999,Samuel2012,OhtaSussman2012} or impulsively accelerated \citep{HanTrygg2001} in another liquid. However, such simulations do not capture the entire fragmentation process since ligament formation and breakup are inherently non-axisymmetric. \cite{Baumann1992} 
have conducted finite volume experiments in immiscible liquid-liquid systems at Weber numbers ranging from $0.3$ to $11000$. $Oh$ is of order one or larger in most of their experiments and only two satisfy $We\geq 100$ and $Oh\ll 1$. \citet{Baumann1992} focus on viscous immiscible vortex rings that form 
at $Re\leq 61$. Instabilities developing on these vortex rings are interpreted as RTI. Several experimental studies of drop breakup in liquid-liquid systems due to shock-induced flows 
have reported drag and breakup time measurements, summarized in \cite{PilchandErdman} and \cite{Gelfand1996}. Among those studies, \cite{PatelTheofanous} show that their breakup time data are consistent with drop piercing by RTI. 
\cite{YangYang} identify a regime where the drop volume grows by turbulent entrainment. 

At large scales, immiscible liquid-liquid plumes \citep{Deguen2011} and immiscible liquid-liquid coaxial jets \citep{Charalampous2008}, at large Weber and Reynolds numbers, are morphologically similar to their miscible equivalents. This suggests that integral models developed for miscible turbulent flows, including 
models of \textit{turbulent thermals} and \textit{vortex rings}, can describe the dynamics of immiscible flows. 


In miscible fluids, a finite buoyant mass 
is called a \textit{thermal} when its impulse originates entirely from the buoyancy force, and a \textit{buoyant vortex ring} when an initial momentum is allowed. As pointed out by Turner (1957, 1964), a thermal can be regarded as a special case of a buoyant vortex ring.
The more general term \textit{vortex ring} refers to a ring-shaped structure formed by closed-loop vorticity lines. 
At high Reynolds numbers, the dynamics of turbulent thermals with small or large density differences \citep{Morton1956,Wang1971,Escudier1973,Baines1984,Thompson2000} and non-buoyant vortex rings \citep{Maxworthy1974} is successfully described using the concept of turbulent entrainment, originally proposed by \cite{Taylor1945} and \cite{Morton1956}, who hypothesized that the rate of growth of a turbulent buoyant mass is proportional to its velocity and surface area. 

The concept of turbulent entrainment has been used to describe the dynamics of two-phase flows in which one phase is dispersed in the other in the form of solid particles \citep{RahimipourWilkinson1992,Bush2003} or air bubbles \citep{Milgram1983,LeitchBaines1989,Bettelini1993}. However, the turbulent entrainment concept applied to immiscible systems that are initially non-dispersed 
has received less attention. It has been used to describe the dynamics of air jets in liquid \citep{Weimer1973,LothFaeth1989,LothFaeth1990}. \cite{EpsteinFauske2001} apply this concept to various liquid-gas and liquid-liquid flows and they develop an erosion model of a liquid drop immersed in a gas or another liquid with an initial velocity lag. They argue that their model is consistent with published data of total breakup time.

In this paper we describe results of a systematic experimental study on the fragmentation of a finite liquid volume into lighter immiscible liquid at low $Oh$, at moderate Reynolds numbers ($Re\geq 10^3$ in most experiments) and for Weber numbers up to $\sim 10^3$. Our main objective is to characterize the different fragmentation regimes in parameter space. Two experimental configurations are used. In the first, the velocity of the released fluid originates entirely from the density difference between the two immiscible fluids 
(immiscible equivalent of thermals). 
In the second, an initial excess in velocity is introduced 
(immiscible equivalent of buoyant vortex rings). 
The experimental apparatus and techniques are described in  $\S2$. 
In $\S3$ we study the early stages of evolution in terms of velocity and deformation. The different fragmentation regimes are characterized in $\S4$ from the study of the subsequent evolution, prior to capillary instabilities and breakup. 
Results on ligament formation and primary breakup are reported in $\S5$. 
At sufficiently high Weber numbers, the flow reaches a turbulent regime whose dynamics are compared, in $\S6$, with predictions from a model based on the concept of turbulent entrainment and on an analogy with miscible thermals and vortex rings. 

\section{Experimental procedure}

\subsection{Experimental set-up}

\begin{figure}
\begin{center}
\includegraphics[width=13.6cm]{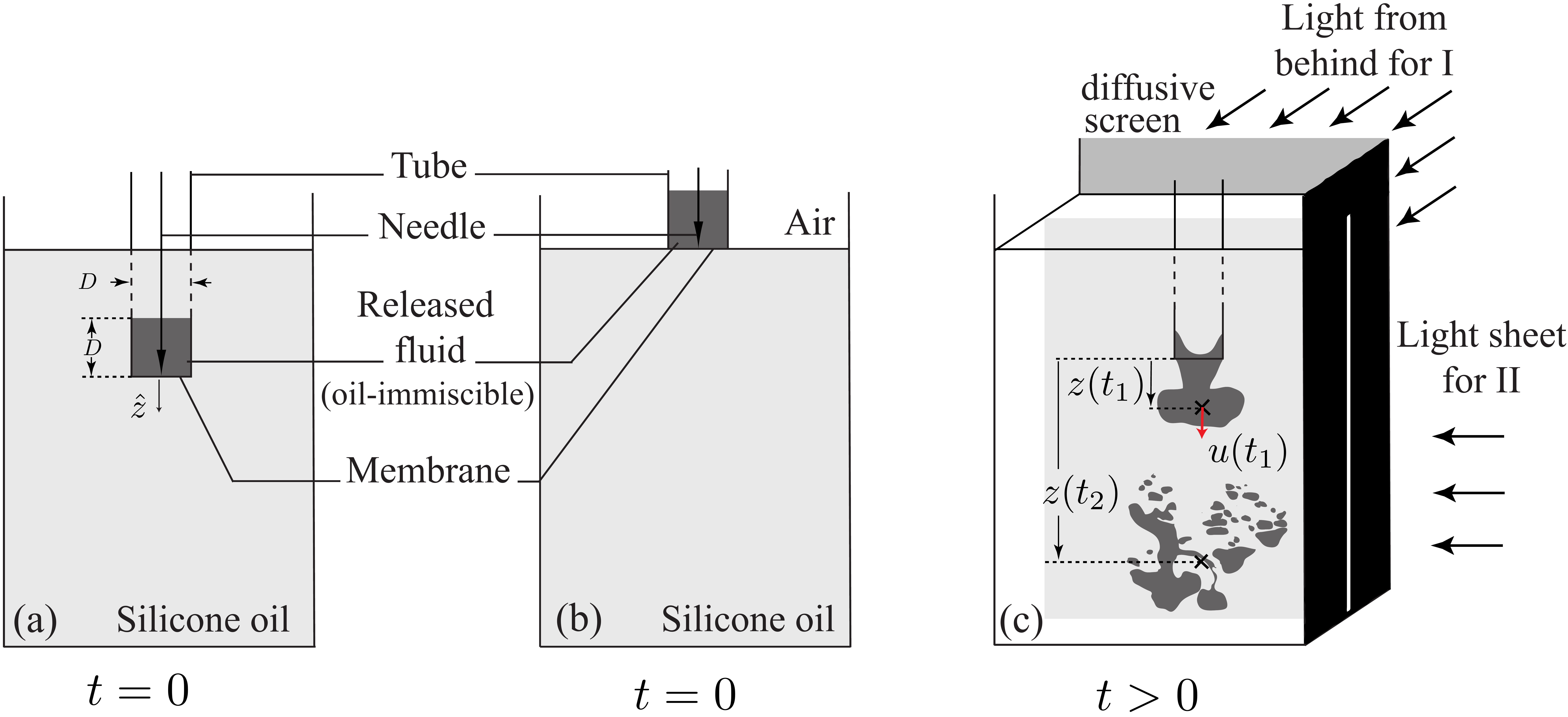}
\end{center}
\caption{Experimental set-up. (a) Side view of the apparatus in the Immersed configuration; (b) side view of the apparatus in the Surface configuration. (c) Visualization techniques and sketch of an experiment with variables measured as a function of time. Symbols I and II refer to backlighting and light-induced fluorescence imaging techniques, respectively. The camera records the side view of the flow depicted in (c). 
}
\label{Dispo}
\end{figure}

The experimental set-up is shown in figures \ref{Dispo}(a,b). A tank of width $25$ cm and height $50$ cm is filled with a low viscosity silicone oil, referred to as the ambient fluid in the following. A denser fluid (detailed below), immiscible in oil, is held in a vertically oriented plastic tube that is closed at the lower extremity by a latex membrane. The denser fluid is released by rupturing the membrane with a needle \color{blue}{inserted into the tube from above, as in the experiments by \cite{Lundgren1992}}. \color{black}The rupture lasts less than $0.04$ s. 

The volume of released fluid $V$ is such that the height of fluid in the tube is equal to the tube internal diameter $D$. Six tubes are used, with $D$ ranging from $1.28$ cm to $7.62$ cm. In the Immersed configuration (figure \ref{Dispo}(a)) the tube is initially immersed in the ambient fluid and it is initially held at the surface of the ambient fluid in the Surface configuration (figure \ref{Dispo}(b)). 



%
%
%


The systematic study has been conducted using backlighting as depicted in figure \ref{Dispo}(c). 
A blue dye (food coloring) is added in the released fluid. The flow is made visible by backward illumination through a diffusive screen and recorded by a color video camera at $24$ frames per second. 
Other flow visualization images are obtained using light-induced fluorescence (figure \ref{Dispo}(c)). The experimental apparatus is illuminated from the side by a light sheet and a fluorescent dye (rhodamine) is added to the released fluid, imaging a cross section of the falling fluid. The light sheet, whose thickness varies from $5$ mm to $7$ mm inside the tank, is produced using a 
flash lamp and a black, opaque screen with a narrow vertical opening of $0.32$ cm. In the following, the backlighting imaging technique is used unless otherwise noted. 
%
%
%
%

In order to vary the density ratio between the ambient and released fluids, different oil-immiscible fluids are used: a mixture of ethanol and water, water, a solution of sodium chloride (NaCl) and a solution of sodium iodide (NaI). Their physical properties are given in Table \ref{Properties}. NaI solution is of particular interest. First, it provides for large density contrasts between the ambient and released fluids, up to the density of silicone oil, without much increase in viscosity. Second, it can be used to match silicone oil refractive index ($n=1.384 \pm 0.006$ at $20^{\circ}$C), which is required to obtain satisfactory images with light-induced fluorescence. Interfacial tension between silicone oil and the released fluid $\sigma$, is measured using a Du No\"{u}y tensiometer.\\
A non-ionic, oil-insoluble surfactant (trade name “Triton X-100”) is added to water and to the NaI solution in several experiments. Equilibrium interfacial tension decreases with surfactant concentration until it reaches the critical micelle concentration, after which it saturates to a constant value. The value given in table \ref{Properties} is used hereafter. 
The highest possible concentration of surfactant $c\approx4$ mL.L$^{-1}$, above which a stable emulsion would be formed in the tank, is used in this study. However, we note that the dynamic interfacial tension may locally be larger than the equilibrium interfacial tension.

In some experiments, water is used in place of silicone oil and a NaCl solution (Table \ref{Properties}) is released. Such experiments are used in \S6 as a reference system.

\begin{table}
\begin{flushleft}
\small
\begin{tabular}{rrrr}
{Fluids} & {Density (kg$\cdot$m$^{-3}$)}&{Viscosity (m$^2\cdot$s)}&{Interfacial tension}\\
{} & {}&{}&{(mN$\cdot$m$^{-1}$)}\\
{} &{} & {}&{}\\
{Silicone oil} &{$820\pm0.2\%$} & {$1.2\times 10^{-6}\pm10\%$}&{}\\
{Ethanol + Water} &{$843.5\pm0.1\%$} & {$2.6\times 10^{-6}\pm10\%$}&{$2.6\pm40\%$}\\
{Water} &{$1000\pm0.05\%$} & {$10^{-6}\pm10\%$}&{$31.2\pm3\%$}\\
{NaCl solution} &{$1175-1192\pm0.06\%$} & {$1.6\times 10^{-6}\pm10\%$}&{$23.3\pm4\%$}\\
{NaI solution} &{$1536-1607\pm0.07\%$} & {$1.3\times 10^{-6}\pm10\%$}&{$17-21\pm10\%$}\\
{Water + Triton X-100} &{$1000\pm0.1\%$} & {$10^{-6}\pm10\%$}&{$3.3\pm15\%$}\\
{NaI sol. + Triton X-100} &{$1260-1578\pm0.06\%$} & {$(1.1-1.3)\times 10^{-6}\pm10\%$}&{$4.4-4.8\pm10\%$}\\
\end{tabular}
\caption{Fluid properties.}
\label{Properties} 
\end{flushleft}
\end{table}

\subsection{Diagnostic techniques}

Preliminary processing (method detailed in appendix \ref{PreliminaryProc}) is first applied to video images (obtained using backlighting) to get binary images. Then, the centroid and velocity of the released fluid are automatically computed. 

We found that the $2D$ centroid obtained from binary images gives too much weight to structures that are located in the rear of the released fluid (membrane of released fluid that remains attached to the tube or wake). Such structures contain a negligible amount of the total released fluid volume whereas they may represent a non-negligible area on a two-dimensional projection.
Instead, we measure a vertical position $z$ that takes into account mass distribution in three dimensions:
\begin{eqnarray}
\label{CM2}
z= \dfrac{\sum\limits_{i,j}z_{i,j}  \log{(I_{i,j}/{I_0}_{i,j})}}{\sum\limits_{i,j} \log{(I_{i,j}/{I_0}_{i,j})}},
\end{eqnarray}

where the pixels $(i,j)$ form the region occupied by the released fluid in the binary image, $z_{i,j}$ is the pixel vertical position, $I_{i,j}$ is the pixel intensity in the original image and ${I_0}_{i,j}$ the pixel intensity in the back field image. The origin $z=0$ corresponds to the lower end of the tube. If the light is monochromatic and the two fluids have the same refractive index, according to the Beer-Lambert law, $z$ is then equal to the depth of the real $3D$ centroid of the released fluid. 
We checked that the dependence of $\log{(I/I_0)}$ on the thickness $l_y$ occupied by released fluid in the direction perpendicular to the image is close to linear if the green band of the image is considered in the range relevant for our experiments. When a nonlinear relationship of the form $l_y=a \log{(I/I_0)}+b \log{(I/I_0)^2}+c\log{(I/I_0)^3}$ with $b=O(a)=O(c)$ is considered, $z$ differs by less than $1\%$ from the value obtained with (\ref{CM2}). Other sources of discrepancy are due to reflection of light on the immiscible interface. According to Fresnel's equations, the reflectivity of the immiscible interface is less than $4 \times 10^{-4}$ in our experiments. Given a rough estimation of the number of droplets and their size, we estimate that the fraction of incident energy reflected on the interface is less than $1\%$ in most experiments and less than $5\%$ in the most turbulent experiments. 
Finally, curvatures of the immiscible interface act as lenses and concentrate light in some portions of the image when the refractive index of the released fluid does not match the refractive index of the ambient fluid. These effects are probably the main source of discrepancies between $z$ and the real 3D centroid. 
From $z$ measurements, we estimate the velocity $u=dz/dt$ of the released fluid as a function of time. 

The MATLAB Image Processing toolbox is used to identify the different connected objects and their equivalent radii in binary images. 
\color{blue}{
First, the holes in a given object are filled to create a simply connected object. Second, this object is divided into two parts by a vertical axis. 
Each part is then rotated $180^\circ$ about the axis to form two half-bodies of revolution. The vertical axis is chosen to pass through the centroid of the resulting $3D$ object. In general, the $3D$ object so constructed consists of two half-bodies of revolution, but in the special case of an image with bilateral symmetry, the $3D$ object would be perfectly axisymmetric. }\color{black} 
The volume of this $3D$ object is given by $V = \sum_{ij} \pi (x_{ij}-\bar{x}) S_{ij} $ where $x_{ij}$ and $S_{ij}$ are the pixel horizontal position and pixel \color{blue}{surface }\color{black} and $\bar{x}$ is the horizontal position \color{blue}{of the centroid of the resulting 3D object}. \color{black} The equivalent radius of the connected object $r$, is defined by $V = \frac{4}{3}\pi r^3$. 

\color{blue}{Uncertainties on $z$ and $r$ measurements are mainly due to their sensitivity to the threshold used to generate binary images (see appendix \ref{PreliminaryProc}). The frame rate of the video camera also affects uncertainties in $u$. Uncertainties on $z$, $r$ and $u$ measurements are typically less than about $5\%$, $5\%$ and $5-10\%$, respectively. In the following (unless otherwise stated), error bars for input dimensionless quantities indicate measurement uncertainties and error bars for output dimensionless quantities correspond to the maximum between measurement uncertainties and standard deviations obtained in series of experiments conducted at the same input parameter values.} \color{black}

\subsection{Input dimensionless numbers}
In Immersed experiments, four input dimensionless numbers govern the dynamics:
\begin{eqnarray}
\label{NDNumbers}
\Bo = \dfrac{\Delta \rho g R^2}{\sigma} \text{,  } \hspace{0.4cm} \Oh = \dfrac{\sqrt{\rho_r} \nu_r}{\sqrt{\sigma R}} \text{,  } \hspace{0.4cm} \DP= \dfrac{\Delta \rho}{\rho_a} \text{,  }\hspace{0.4cm} \dfrac{\nu_r}{\nu_a}.
\end{eqnarray}
Here $\Bo$ is the Bond number, $\Oh$ the Ohnesorge number, $\Delta \rho$ is the density difference between the ambient and released fluids, $g$ the acceleration due to gravity, $R$ the equivalent spherical  radius of the released fluid, 
$\nu$ kinematic viscosity, $\rho$ density. The subscript $a$ and $r$ denote the ambient and released fluid, respectively. $Bo$ measures the importance of the buoyancy force versus interfacial forces. 
In Surface experiments additional dimensionless numbers are introduced since the released fluid is initially surrounded by air. 
We are interested in the fragmentation of released fluid in oil and we do not consider interfacial effects involving air. The density and viscosity ratios between air and silicone oil should be added to the above set of dimensionless numbers, however their values remain constant in all the experiments.

Experiments have been conducted for $24$ different sets of input dimensionless numbers in the Immersed configuration and $30$ sets in the Surface configuration. $\Bo$ and $\DP$ lie in the range $\sim4-1430$ and $\sim0.029-0.96$ respectively, $\Oh$ varies from $\sim 10^{-3}$ to $\sim 10^{-2}$ and $\nu_r/\nu_a$ from $0.8$ to $2.2$. Since $\Oh\ll1$ we expect viscosity to have little influence on the fragmentation regime in agreement with previous studies on drop fragmentation (\cite{Hinze1955}, \cite{PilchandErdman} and reviews in \cite{Gelfand1996} and \cite{Guildenbecher2009}). 
In this study, we thus concentrate on the effects of $Bo$ and $P$, which are independent of viscosity. 

\section{Early stages of evolution: post-release conditions}
In this section we study the velocity and deformation of the released fluid at a short distance from the tube ($z\lesssim2R$) for the two experimental configurations used.


\subsection{Weber number scaling: post-release velocity}



The definitions of the Weber and Reynolds numbers involve a characteristic velocity $U$ such that 
\begin{eqnarray}
\Web=\frac{\rho_r U^2 R}{\sigma}\hspace{0.2cm}; \hspace{1cm} \color{blue}{\Rey = \frac{UR}{\nu_a}}.
 \label{We}
\end{eqnarray}
In this subsection, we define $U$ and extract a scaling law for $\Web$ as a function of the input dimensionless numbers $\Bo$ and $\DP$. 


The characteristic velocity classically used at high Reynolds numbers is the terminal velocity, a balance between buoyancy and form drag forces, which gives $U \propto \sqrt{gPR}$. However this scaling is not appropriate for our experiments since fragmentation processes start before the released fluid has reached its terminal velocity (expected between $10-20 R$). In addition, at fixed $Bo$ and $\DP$, the vertical velocity at short distances $z$ is larger in the Surface configuration than in the Immersed configuration. This results from the buoyancy force being initially larger in the Surface configuration since it involves the density difference between the released fluid and the air, rather than $\Delta \rho$. This velocity excess is \color{blue}{not} \color{black} accounted for by the terminal velocity scaling which predicts the same characteristic velocity in both configurations.

Another natural scaling, which is adopted here, emerges from a balance between the rate of change in released fluid momentum and buoyancy forces, by assuming that a given portion of the mechanical work generated by buoyancy forces (potential energy) is converted into kinetic energy of the released fluid during its fall.


\subsubsection{Immersed configuration}
In the Immersed configuration, this scaling takes the form
\begin{eqnarray}
 \frac{1}{2} \rho_r u^2 \propto \Delta \rho g (z+D/2),
 \label{EpEc}
\end{eqnarray}
where the distance to the tube end $z$ is initially equal to $-D/2$. Scaling (\ref{EpEc}) implies that the characteristic velocity $U$ should be defined at a given distance from the tube $Z$. 
The choice of $Z$ is partly arbitrary but two conditions have to be met: the released fluid is entirely off the tube at $z=Z$ and drop formation has not yet started. $Z=2R$ satisfies both conditions in our experiments.

\begin{figure}
\begin{center}
\includegraphics[width=13.4cm]{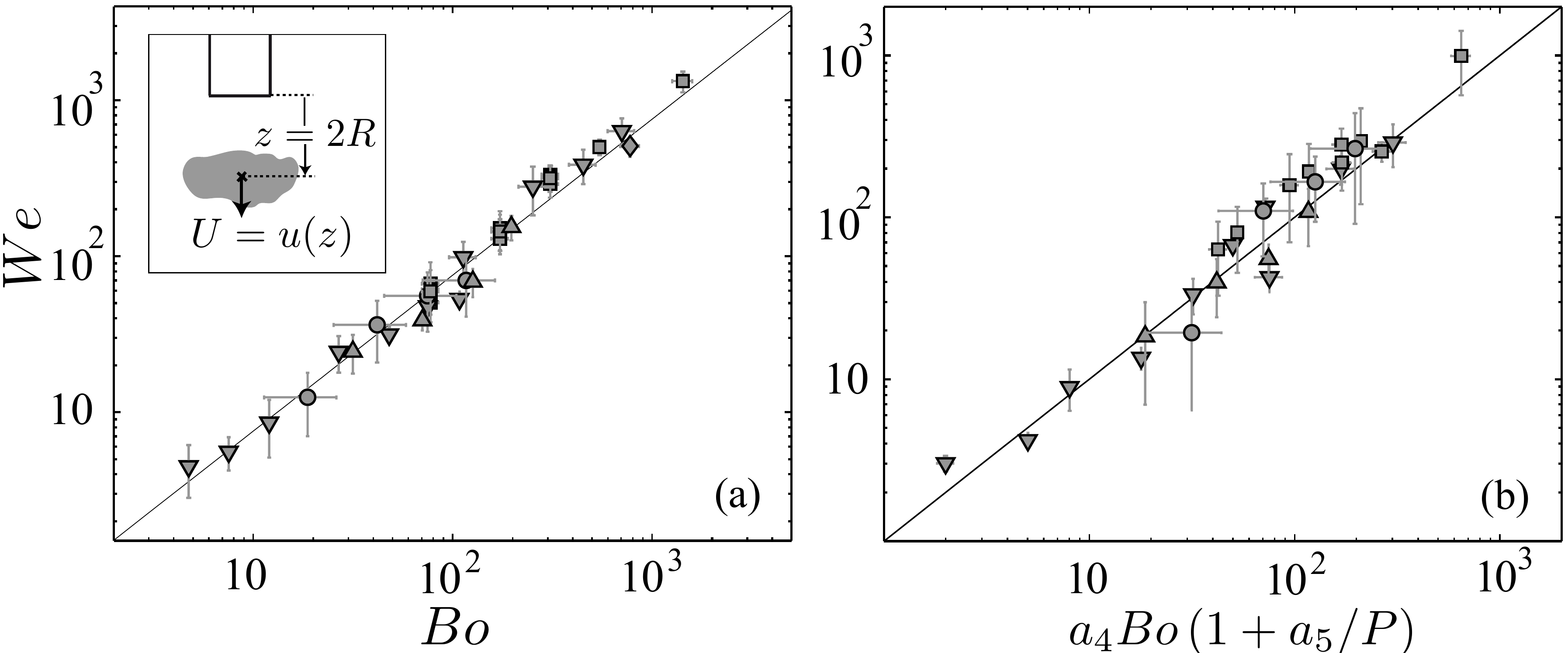}
\end{center}
\caption["Short" caption without tikz code]{(a) Weber number versus Bond number in the Immersed configuration using \hbox{$U=u(z=2R)$} as illustrated in the insert. The least squares best fit (\ref{BFIm}) is shown by the black line. (b) Weber number measured in Surface experiments with $U=u(z=2R)$ versus Weber number predicted by the least squares best fit (\ref{BFS})-(\ref{BFS2}). 
\greysquare{black} , $0.82\leq \DP \leq 0.96$; \greydiamond{black}  , $\DP \approx 0.54$;  \greytriangleup{black}   , $\DP \approx 0.43$; \greytriangle{black}   , $\DP \approx 0.22$; \greycircle{black} , $\DP \approx 0.03$.  
}
\label{U_Tot}
\end{figure}

The potential/kinetic energy balance (\ref{EpEc}) implies $\Web \propto \Bo$, which is in agreement with the experimental data shown in figure \ref{U_Tot}(a). 
We obtain the following least squares best fit:
\begin{eqnarray}
\Web = a_1\Bo \hspace{0.2cm} \text{, } \hspace{0.2cm} a_1 = 0.76\pm0.04,
\label{BFIm}
\end{eqnarray}
\color{blue}{with a mean deviation of $17\%$ relative to the experimental data. 
A Monte Carlo method of error propagation was used to estimate the standard deviation of $a_1$ from $50$ synthetic data sets formed of pseudorandom numbers within the experimental errorbars, one number for each data point in figure \ref{U_Tot}(a). } \color{black}

\subsubsection{Surface configuration}
In the Surface configuration it is not straightforward to estimate the mechanical work generated by buoyancy forces. For example, the buoyancy force involves the density difference with the surrounding air $\rho_r-\rho_{air}\approx \rho_r$ at initial times, but once the fluid is entirely immersed in the ambient fluid, it depends only on $\Delta \rho$. Assuming that the mechanical work generated by buoyancy forces can be written as a sum of  
two independent terms, originating from the former contributions, and assuming that a portion of this work is converted into kinetic energy, we obtain 
\begin{eqnarray}
\frac{1}{2}\rho_r U^2 = a_2 \Delta \rho g R + a_3 \rho_r g R,
\label{EpEc_surf}
\end{eqnarray}
 where $a_2$ and $a_3$ are two constants to be fitted. In terms of dimensionless numbers (\ref{EpEc_surf}) amounts to 
\begin{eqnarray}
\Web=a_4 \Bo \left(1+\frac{a_5}{\DP} \right),
\label{BFS}
\end{eqnarray}
where $a_4=a_2+a_3$ and $a_5=a_3/(a_2+a_3)$.
The experimental results are shown in figure \ref{U_Tot}(b). We find the following least squares best-fit values: 
\begin{eqnarray}
a_4=0.52\pm0.07 \hspace{0.2cm}\text{    and     } \hspace{0.2cm}a_5=0.07 \pm 0.03, 
\label{BFS2}
\end{eqnarray}
\color{blue}{with a mean deviation of $22\%$ relative to the experimental data. Standard deviations in $a_4$ and $a_5$ 
are estimated using the same method as for $a_1$ in (\ref{BFIm}).} \color{black}



We have 
found scaling laws for $\Web$ as a function of the input dimensionless numbers which fit reasonably well with the experimental data. As a consequence, $\Bo$ and $\Web$ can substitute for each other and the physical processes can be studied alternatively in a $(\Bo,\DP)$ or $(\Web, \DP)$ diagram. We will mainly concentrate on the $(\Web, \DP)$ diagram 
since, as shown in \S4, it is well-suited for comparisons between fragmentation regimes in the Immersed and Surface configurations. 
$\Rey$ varies from $\sim \color{blue}{500}$ to $\sim  10^4$ in our experiments, with $\Rey\geq 10^3$ in a large majority of experiments ($\color{blue}{90\%}$). We do not concentrate on the effect of $\Rey$ since, as we have already argued in \S2, viscosity is expected to have little influence on the fragmentation regime. This is confirmed by estimations of the capillary number $Ca = \nu_r \rho_r U/\sigma$, which measures the ratio of viscous forces to interfacial forces: $Ca$ remains much smaller than $1$ in our experiments (in the range $\sim0.005-0.1$). 

\subsection{Early deformations and destabilizations: post-release shape}


Once the released fluid exits the tube, it starts to deform and change shape. A wide variety of shapes is observed directly after the release ($z\lesssim 2R$) as illustrated in figures \ref{RT}(a,b,c) and figure \ref{VR}. The present section aims at understanding the physical mechanisms involved. 


\subsubsection{Immersed configuration}

\begin{figure}
\begin{center}
\includegraphics[width=13cm]{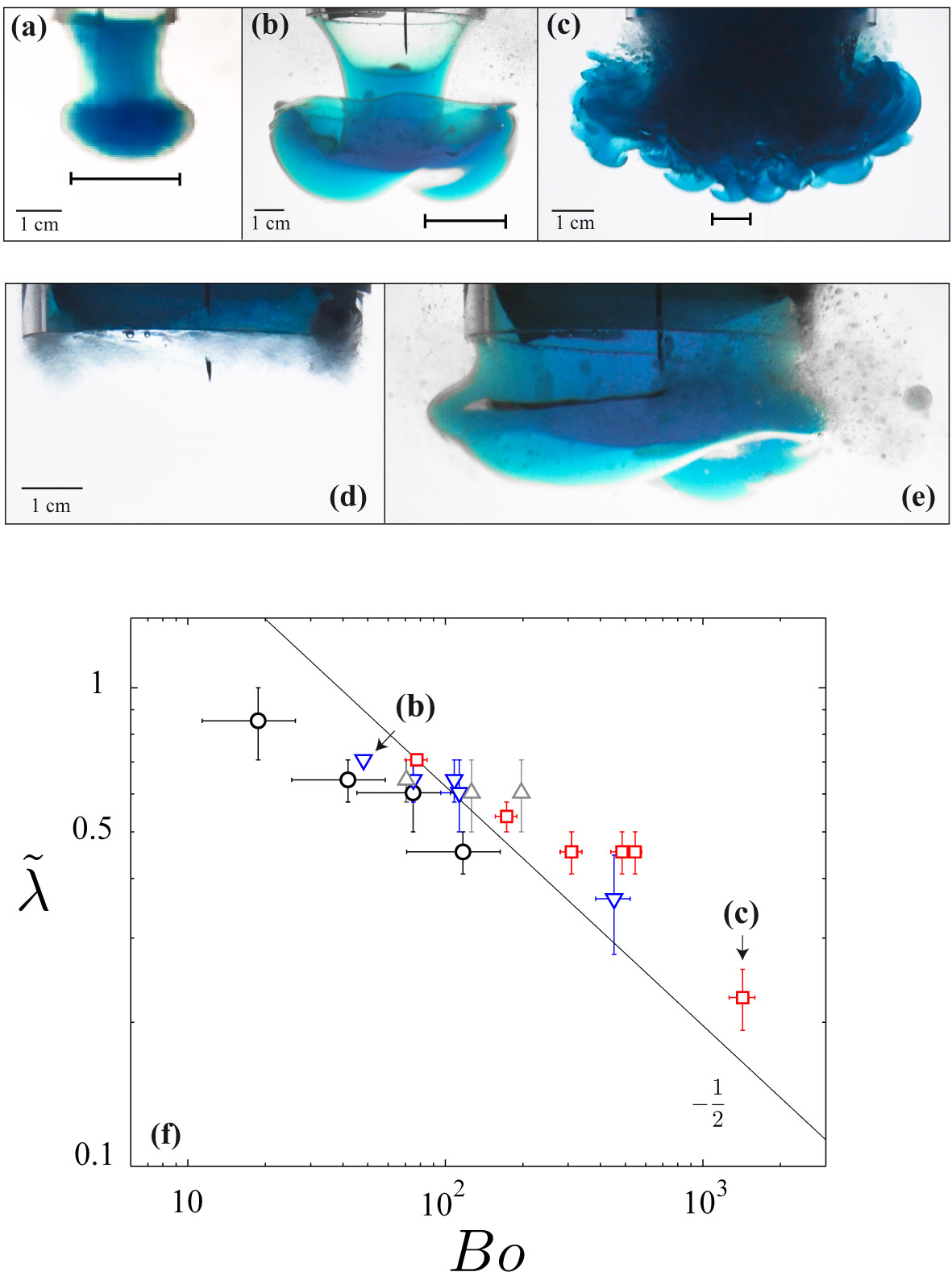}
\end{center}
\caption["Short" caption without tikz code]{(a,b,c) Early deformations of the released fluid in Immersed experiments for $z\lesssim 2R$: at (a) $\Web\approx 10$, (b) $\Web\approx \color{blue}{30} \color{black} $, and (c) $\Web\approx 1.5\times 10^{3}$. The black mark in (a) indicates the critical wavelength of RTI deduced from equation (\ref{LambdacND}) whereas black marks in (b) and (c) indicate the most amplified wavelength of RTI deduced from equation (\ref{LambdaND}). 
\color{blue}{(d,e) Deformations in the aftermath of the membrane rupture, same experiment as in (b) at two preceding times, separated by about $0.2$ s. 
}\color{black} (f) Estimated dimensionless wavelength $\tilde{\lambda}=1/\sqrt{n}$ as a function of $Bo$ in Immersed experiments where $n$ is the number of mushroom-shaped structures. \bicoloursquare{white}{red} , $0.87\leq\DP \leq0.96$;  \bicolourtriangleup{white}{gray}   , $\DP \approx 0.43$; \bicolourtriangle{white}{blue}   , $\DP \approx 0.22$; \bicolourcircle{white}{black} , $\DP \approx 0.03$. The black curve gives the most amplified wavelength predicted by equation (\ref{LambdaND}). 
Symbols (b) and (c) denote results obtained from the experiments shown in (b) and (c). \color{blue}{The allowable slope is within the range: $(-0.3,-0.2)$ when considering all the experiments, $(-0.45,-0.25)$ for 
$0.87\leq\DP \leq0.96$, $(-0.4,-0.2)$ for $\DP \approx 0.43$, $(-0.3,-0.15)$ for $\DP \approx 0.22$ and $(-0.5,-0.1)$ for $\DP \approx 0.03$.} \color{black}}
\label{RT}
\end{figure}

Figures \ref{RT}(a,b,c,\color{blue}{d,e)} \color{black} illustrate the initial deformations of the released fluid in the Immersed configuration at different Weber numbers. At $\Web\approx 10$ (a) the released fluid flattens into a pancake shape due to dynamic pressure forces while non-axisymmetric perturbations are damped. 
At $\Web\approx30$ (b,d,e) non-axisymmetric perturbations grow and at $\Web\approx 1.5 \times 10^3$ (c) these non-axisymmetric structures develop a mushroom shape, which is morphologically similar to Rayleigh-Taylor instabilities (RTI).

We first compare these with the classical inviscid analysis of the Rayleigh-Taylor instability of a horizontally unbounded interface between two immiscible fluids \citep{BellmanPennington1954,ChandraBook}. Choosing a coordinate system that 
moves with the released fluid, 
the governing equations are left unchanged if $g$ is replaced by $g-du/dt$. The uncertainties on $d u/dt$ measurements are too large for any scaling law to be extracted but we estimate that its maximum value is of order $1-2$ m$\cdot$s$^{-2}$ and therefore, as a first order approximation, we neglect $du /dt$ with respect to $g$. 
Then, in the case of vertically unbounded layers, the growth rate $\gamma$ of small perturbations is given by
\begin{eqnarray}
\gamma = \sqrt{\frac{\Delta \rho}{\rho_a+\rho_r}g k - \frac{\sigma}{\rho_a+\rho_r} k^3},
\label{TauxCroiss}
\end{eqnarray}
where $k$ is the wavenumber of the disturbances at the interface. The most amplified and critical wavenumbers are respectively given by 
\begin{eqnarray}
\label{km}
k_m = \sqrt{\dfrac{\Delta \rho g}{3\sigma}},
\\
k_c = \sqrt{\dfrac{\Delta \rho g}{\sigma}}.
\label{kc}
\end{eqnarray}

It can be shown that including viscous effects \citep[see equation (113) from][Chap. X]{ChandraBook} has little effect on the value of $k_m$ and $k_c$ at parameter values relevant for our experiments. In terms of dimensionless wavelength $\tilde{\lambda} = 2 \pi / k D$ equations (\ref{km}) and (\ref{kc}) take the form

\begin{eqnarray}
\label{LambdaND}
\tilde{\lambda}_m =   \dfrac{2\pi \sqrt{3}}{\sqrt{\Bo}}\dfrac{R}{D},
\\
\tilde{\lambda}_c =  \dfrac{2\pi}{\sqrt{\Bo}} \dfrac{R}{D}.
\label{LambdacND}
\end{eqnarray}

The most amplified wavelength predicted by (\ref{LambdaND}) matches the size of non-axisymmetric \color{black}structures in figures \ref{RT}(b,c). In figure \ref{RT}(a) the predicted critical wavelength $\tilde{\lambda}_c D$ is close to the tube diameter which explains why RTI do not develop at the front of the released fluid. The number of mushroom-shaped structures $n$ is evaluated in our experiments from backlighting images (e.g. figures \ref{RT}(b,c)) 
and a characteristic wavelength is estimated by $\tilde{\lambda}=1/\sqrt{n}$. \color{blue}{When the size of the mushroom-shaped structures is of the same order of magnitude as the tube diameter (e.g. figure \ref{RT}(b)), the number of structures that can be hidden on the back side is small (included in the error bar). On the contrary, when the size of the structures is much smaller than the tube diameter, the lower bound on $n$ is given by the number of mushroom-shaped structures that can be counted on backlighting images (e.g. figures \ref{RT}(c)), and the upper bound is estimated by assuming that the back side hides this same number. 
Alternatively, we could have estimated a characteristic wavelength by measuring the mean width for the mushroom-shaped structures. However, the latter method would be more subjective, because the instabilities are three-dimensional and their width depends on the vertical section considered, and because the width of the mushroom-shaped structures varies with time in a single experiment while the number of waves remains the same.} \color{black}
\color{blue}{Figure \ref{RT}(f) shows that the resulting absolute values for $\tilde{\lambda}$ are in the same order of magnitude as the theoretical predictions (\ref{LambdaND}). 
However, the allowable slope in figure \ref{RT}(f) is more shallow than the slope predicted by (\ref{LambdaND}). This discrepancy is somewhat reduced at a fixed $\DP$ value (figure \ref{RT}(f)), though the prefactors vary with $\DP$, in a way not predicted by equation (\ref{LambdaND}). 

Such differences between the experimental wavelength and equation (\ref{LambdaND}) may be due to the simplicity of the above model, in which several effects discussed 
below have been neglected, or to 
initial vorticity perturbations due to the release mechanism.}  \color{black} 
\color{blue}{
Indeed, the membrane retraction induces an emulsified layer near the immiscible interface 
(figure \ref{RT}(d)); the layer forms in less than $0.04$ s and is then swept around the sides of the released fluid during the fall on the advective time scale (figure \ref{RT}(e)). 
This emulsified layer probably results from the flapping of the membrane combined with wake instabilities 
that involve shear boundary layers of opposite vorticity generated on the upper and lower surfaces of the retracting membrane. 
The membrane rupture might then affect the pattern of initial disturbances at the interface, however, showing $\Web/\Bo$ is constant even for small $\Bo$ (figure \ref{U_Tot}(a)) indicates that 
the circulation generated during the membrane rupture does not contribute significantly to the total circulation of the released fluid.
}\color{black}  

\begin{figure}
\begin{center}
\includegraphics[width=12cm]{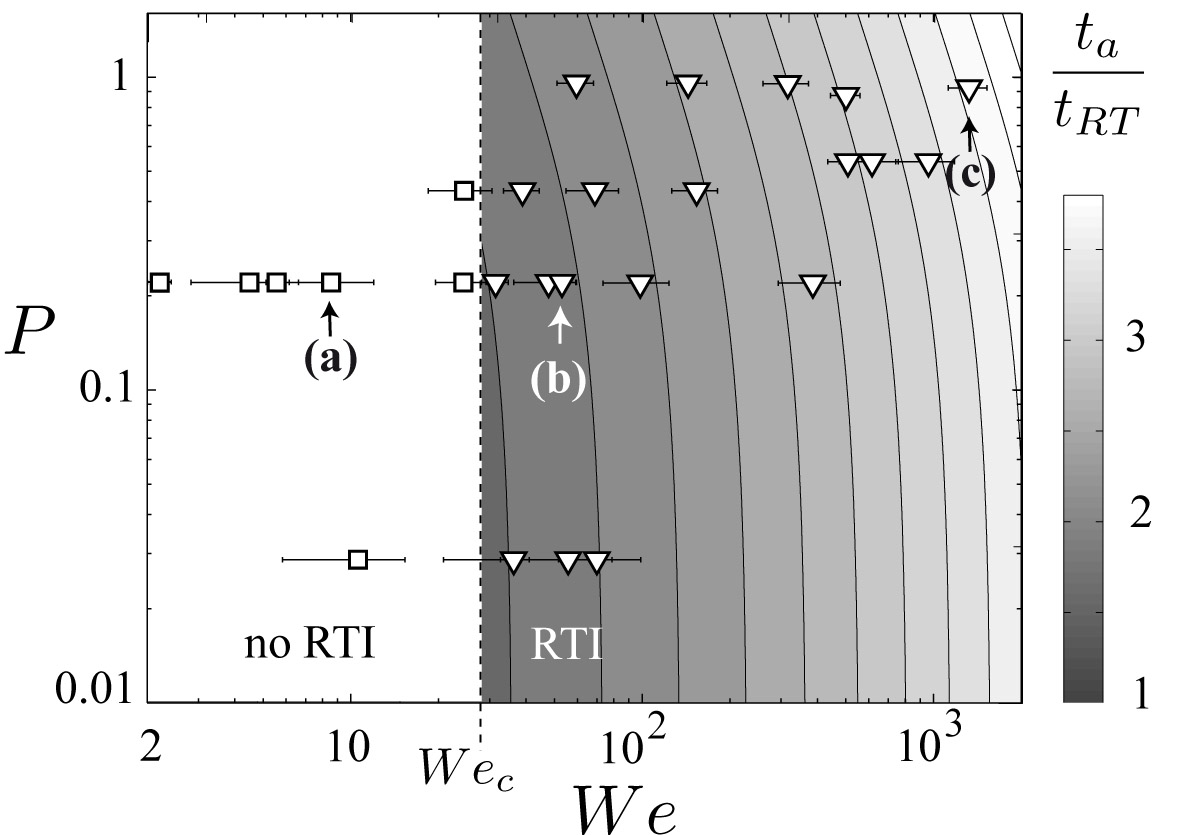}
\end{center}
\caption["Short" caption without tikz code]{$(\Web, \DP)$ diagram of the early stage deformation in the Immersed configuration. \bicoloursquare{white}{black} , axisymmetric deformations (no RTI); \bicolourtriangle{white}{black}  , RTI. \color{blue}{Dashed line: tentative boundary}\color{black}. The ratio of the advective timescale to the RTI timescale $t_a/t_{RT}$ is color-coded in regions where RTI are found, at $\Web\geq We_c$. Symbols (a),(b),(c) denote the experiments shown in \color{blue} figures \ref{RT}(a,b,c) \color{black}. }
\label{DP_Init_Im}
\end{figure}

\begin{figure}
\begin{center}
\includegraphics[width=14cm]{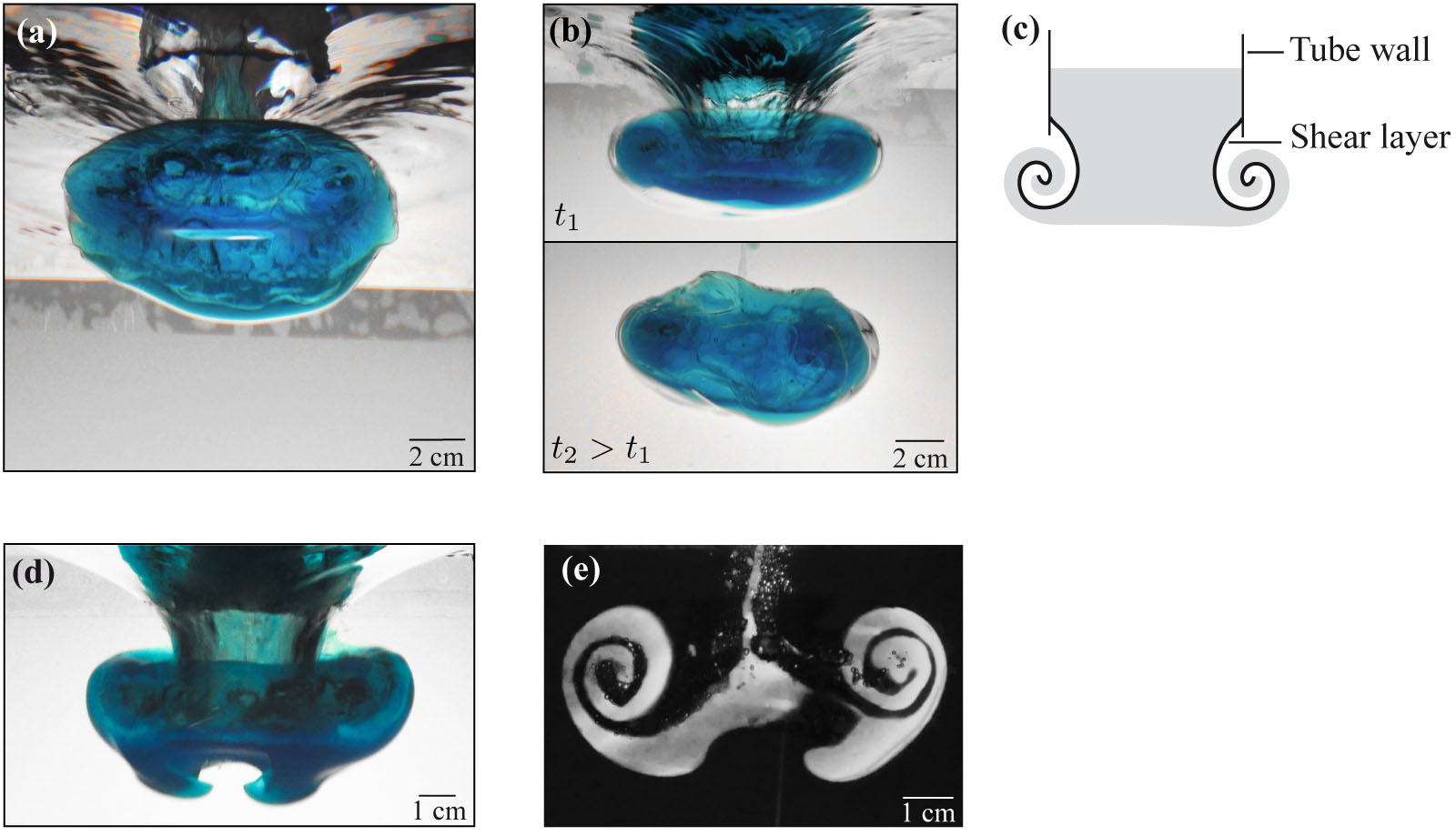}
\end{center}
\caption{(a,b,d,e) Early deformations in the Surface configuration. (a,b) $\Web\approx 100$, $\DP\approx 0.22$; $z\lesssim R/4$ and $z\approx 2.5 R$ in, respectively, the first and second snapshot of (b); (d) $\Web\approx 300$, $\DP\approx 0.96$, $z\approx R$; (e) obtained with light-induced fluorescence, $\Web\approx 200$, $\DP\approx 0.54$, $z\approx 2R$. (c) Schematic representation of the generation of a shear layer and its roll-up to form a vortex ring.}

\label{VR}
\end{figure}

A regime diagram of the initial deformations is shown in figure \ref{DP_Init_Im}. Non-axisymmetric perturbations emerge at $\Web_c$, which is located between $\sim 20$ and $\sim30$ given the uncertainties on $\Web$. From equation (\ref{LambdacND}) we estimate that the number of mushroom-shaped structures is equal to $n_c=2$ when the Bond number is equal to $Bo_c=8\pi^2(R/D)^2$, assuming $\tilde{\lambda}_c=1/\sqrt{n_c}$. Applying the experimental scaling (\ref{BFIm}), we find $\Web_c\approx 20\pm 1$, which is \color{blue}{broadly} \color{black} consistent with the experimental results. \color{blue}{Overall, the early non-axisymmetric perturbations emerging in our immersed experiments are in reasonable agreement with RTI.} \color{black} 
\color{blue}{Several effects have been neglected in the above analysis of RTI.} \color{black} First, the fluid layers are not vertically unbounded. It \color{blue}can be \color{black} shown that this effect has a secondary impact in the linear regime given the value of $\tilde{\lambda}/D$ in our experiments and, equations (\ref{TauxCroiss}), (\ref{km}) and (\ref{LambdaND}) remain valid at first order. 
Second, the released fluid is confined in the horizontal direction. 
 \citet{JacobsCatton1988a} have shown that geometry does not enter in the linear stability analysis of a fluid layer confined in a circular container of diameter $D$ and overlying a gas layer. 
 A similar result holds in the case of two fluid layers 
 so that equation (\ref{TauxCroiss}) remains valid. The circular geometry quantizes the possible values of the wavenumber $k$: $kD/2$ has to be a zero of the Bessel functions of the first kind. 
However this effect has little impact on our conclusions since the characteristic dimensionless size of the most amplified waves follows the same general trend $ \propto 1/\sqrt{\Bo}$ as in (\ref{LambdaND}). 
Finally, in the analysis leading to equation (\ref{TauxCroiss}), the undisturbed state is at rest in the moving coordinate system. In our experiments an axisymmetric basic flow develops during the fall of the released fluid, advecting the growing RTI from unstable regions at the front to stable regions at the rim. 
These effects have been examined in previous studies on the 3D instability of bubbles rising through liquid. 
\citet{Grace1978} proposed a semi-empirical model based on the idea that breakup occurs if the characteristic timescale for RTI growth $t_{RT}=1/\gamma_m$, where $\gamma_m$ is the maximum growth rate, is small enough compared to the time available for growth, i.e. the advective timescale, $t_a$. \citet{Batchelor1987} improved this model by including a basic flow, assumed to be axisymmetric and irrotational, in the stability analysis. He showed that the contractional motion in the direction normal to the interface tends to decrease the amplitude of a disturbance while its wavelength increases exponentially due to the extensional motion parallel to the interface. Because of the latter effect, disturbances do not grow exponentially with a constant growth rate. 
Similar effects are expected in our experiments, but the axisymmetric basic flow is inherently time-dependent, causing an increase in complexity. 
For this reason, we treat advection and RTI as if they were two independent mechanisms 
as in \citet{Grace1978}. 

As a first approximation we use $t_{RT}=1/\gamma(k_m)$ where $\gamma$ and $k_m$ are given by equations (\ref{TauxCroiss}) and (\ref{km}). 
Taking $R/U$ as a characteristic advective timescale $t_a$ we obtain

\begin{eqnarray}
\label{taRT}
\dfrac{t_{a}}{t_{RT}} = d_1\dfrac{\Bo^{3/4}}{\Web^{1/2}} \sqrt{\frac{1+\DP}{2+\DP}},
\end{eqnarray}

where $d_1=\sqrt{2/3^{3/2}}$. Making use of the experimental scaling (\ref{BFIm}), equation (\ref{taRT}) takes the form (used in figure \ref{DP_Init_Im})

\begin{eqnarray}
\label{taRT_Im}
\dfrac{t_{a}}{t_{RT}} = \dfrac{d_1\Web^{1/4}}{a_1^{3/4}} \sqrt{\frac{1+\DP}{2+\DP}}, \hspace{0.3cm} a_1 = 0.76\pm0.04.
\end{eqnarray}


According to (\ref{taRT_Im}), $t_{a}/t_{RT}$ varies weakly with $\DP$ (figure \ref{DP_Init_Im}), which is consistent with no observed change in the deformation regime when varying $\DP$ at a fixed $\Web$ value.  Equation (\ref{taRT_Im}) predicts that RTI remain the dominant mechanisms when $\Web$ increases, which is also consistent with experimental observations (figure \ref{DP_Init_Im}). Close to $\Web=\Web_c$, $t_a/t_{RT}\sim1$, indicating that the effect of advection of RTI by the basic flow is probably significant. This may be responsible for a short delay in the emergence of RTI ($We_c$ in figure \ref{DP_Init_Im}) compared to the critical value $\Web\approx 20$ predicted from equation (\ref{LambdacND}). 


\subsubsection{Surface configuration}

\begin{figure}
\begin{center}
\includegraphics[width=12cm]{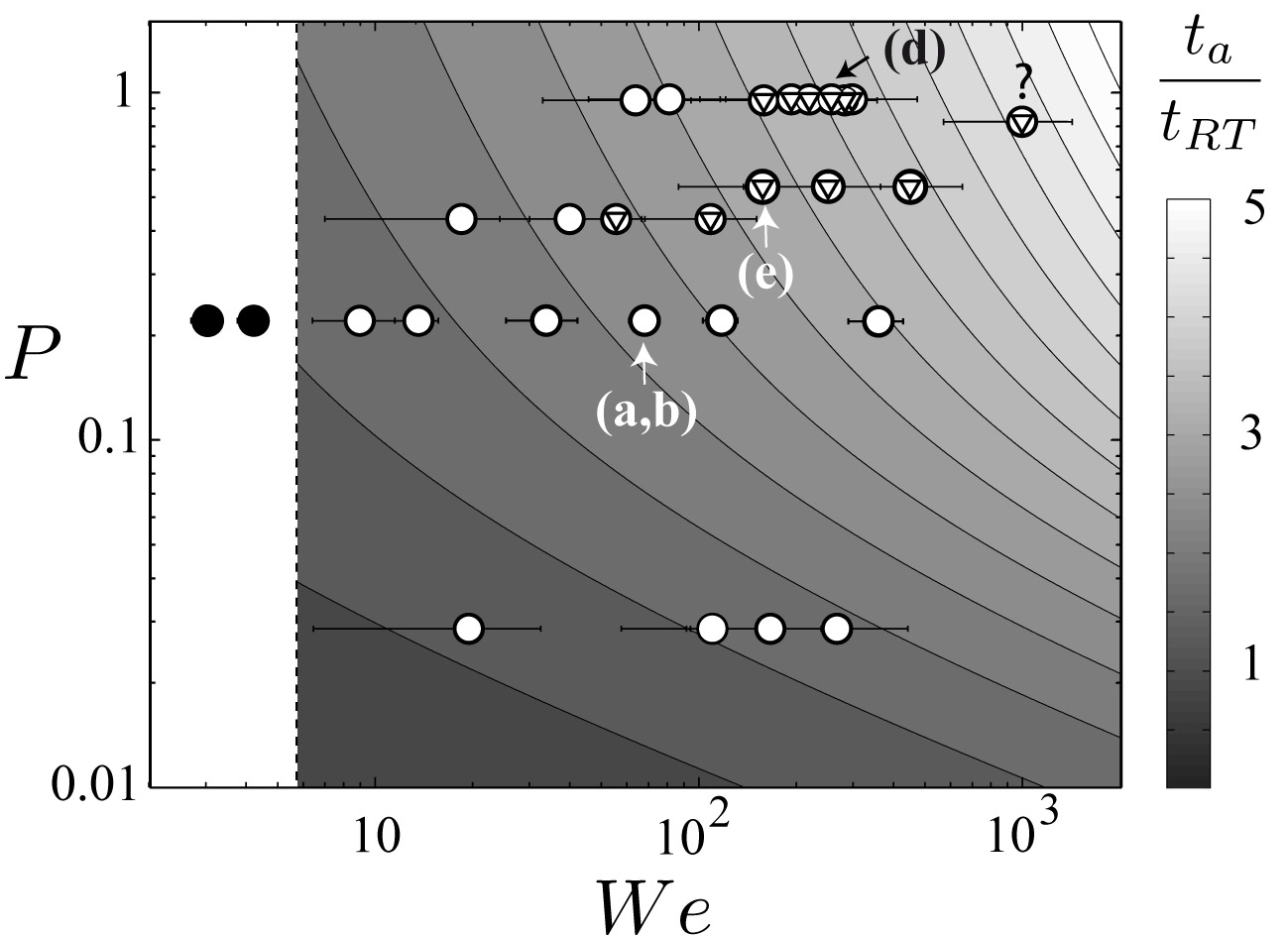}
\end{center}
\caption["Short" caption without tikz code]{\color{black}$(\Web, \DP)$ diagram of the first deformations in the Surface configuration. \bicolourcircle{black}{black} , oblate drops; \bicolourcircle{white}{black} , vortex rings; \bicolourtrianglec{black}{white}, mushroom-shaped structures, typical of RTI, are observed at the front of a vortex ring.  \color{blue}{Dashed line: tentative boundary}\color{black}. The question mark denotes an experiment in which no clear visualization of mushroom-shaped structures was captured but waves of characteristic size consistent with the predicted wavelength for RTI are observed. The ratio of the advective timescale to the RTI timescale $t_a/t_{RT}$ is color-coded \color{blue}{in regions where a vortex ring forms}\color{black}. 
\color{blue}{Symbols (a),(b),(d),(e) denote the experiments shown in \color{blue} figures \ref{VR}(a,b,d,e) \color{black}.}
}
\label{DPInit_surf}
\end{figure}

In the Surface experiments, when $\Web \gtrsim \color{blue}{6}$, a vortex ring forms at the tube end, as a result of the roll-up of a shear layer generated at the tube wall during the release (figures \ref{VR}(a,b,c)). \color{blue}{
Perturbations caused by the membrane rupture have no strong effect on the shear layer roll-up since nearly axisymmetric rings are formed (figures \ref{VR}(a,b)), 
in agreement with the observations by \cite{BondJohari2010} for miscible vortex rings.} \color{black} Contraction of the initial ring's diameter is observed in most experiments at $z\sim2R-3R$ (figure \ref{VR}(b), see also supplementary video $2$). A decrease in the ring diameter after its formation has already been reported in experiments \citep{Didden1979} and numerical simulations \citep{Nitsche1994}, and is due to the influence of the tube orifice \color{black}\citep{Didden1979,Sheffield1977} or a secondary vortex of opposite circulation formed on the tube end \citep{Didden1979}. In our experiments, the release process generates a strong wave at the surface of the tank, causing penetration of ambient fluid into the tube after the release is completed, and possibly responsible for the generation of a secondary vortex. Mushroom-shaped structures are observed at the front of the vortex ring in experiments located at the highest $\DP$ values (figures \ref{VR}(d,e) and figure \ref{DPInit_surf}). 




Using the same argument as in \S3.1.2, we hypothesize that RTI emerge in Surface experiments when the characteristic time for disturbance growth $t_{RT}$ is small compared to the advective timescale $t_a$. In previous experimental studies of non-buoyant vortex rings generated by a piston \citep{Gharib1998} and in the numerical study of the roll-up of a vortex sheet \citep{Moore1974}, it has been shown that the characteristic timescale for the formation of the vortex ring is few advective times, suggesting that the competition between the growth of perturbations at the front and their advection by the flow is a competition between disturbances growth and the roll-up of the shear layer.

Once the released fluid is entirely immersed, the buoyancy force becomes the same as in an equivalent Immersed experiment. At this stage, $du / dt$ is smaller than $\sim 0.2g$ 
and, with the same assumptions and limitations as in \S$3.2.1$, $t_a/t_{RT}$ is given by equation (\ref{taRT}). Making use of scaling (\ref{BFS}), we obtain 
\begin{eqnarray}
\label{taRT_Surf}
\dfrac{t_{a}}{t_{RT}} = \dfrac{d_1\Web^{1/4}}{a_4^{3/4}\left( 1+ a_5/\DP\right)^{3/4}} \sqrt{\frac{1+\DP}{2+\DP}},
\end{eqnarray}
where $a_4=0.52\pm0.07$ and $a_5=0.07\pm0.03$. 
Equation (\ref{taRT_Surf}) predicts that, contrary to the Immersed configuration, $t_a/t_{RT}$ strongly depends on $\DP$, which explains why the deformation regime changes in Surface experiments when varying $\DP$ at a fixed $\Web$ (figure \ref{DPInit_surf}). Mushroom-shaped structures are found at the largest $\DP$ and $\Web$ values, in regions where $t_a/t_{RT}$ reaches its highest values, consistent with the hypothesis that these structures result from RTI. 

\color{blue}
We note that, in Immersed experiments, a shear layer also starts to roll-up during the release, below the tube edge (figure \ref{RT}(c)), 
for sufficiently large $\Web$ ($\gtrsim$ 50). 
However, the roll-up of the shear layer is interrupted by 
RTI 
before any vortex ring can be formed. 
In the Surface configuration, $d u/dt$ reaches $0.4g$ in the aftermath of the membrane rupture, a value larger than in the Immersed configuration because the former case initially involves a density contrast $\rho_r-\rho_{air}\geq \Delta \rho$. First, this implies that the initial effective acceleration $a=g-du/dt$ is smaller in Surface experiments, reducing the growth rate of RTI by a factor of about $2$. Second, larger velocities are reached during the early stages of the fall and the total circulation of the vortex sheet $\Gamma$ is larger in the Surface configuration, which tends to decrease the vortex sheet roll-up time. Such effects qualitatively explain why the vortex ring roll-up can be completed in Surface experiments and not in Immersed experiments. 
\color{black}
\section{Subsequent evolution : characterization of fragmentation regimes}

\begin{figure}
\begin{center}
\includegraphics[width=9cm]{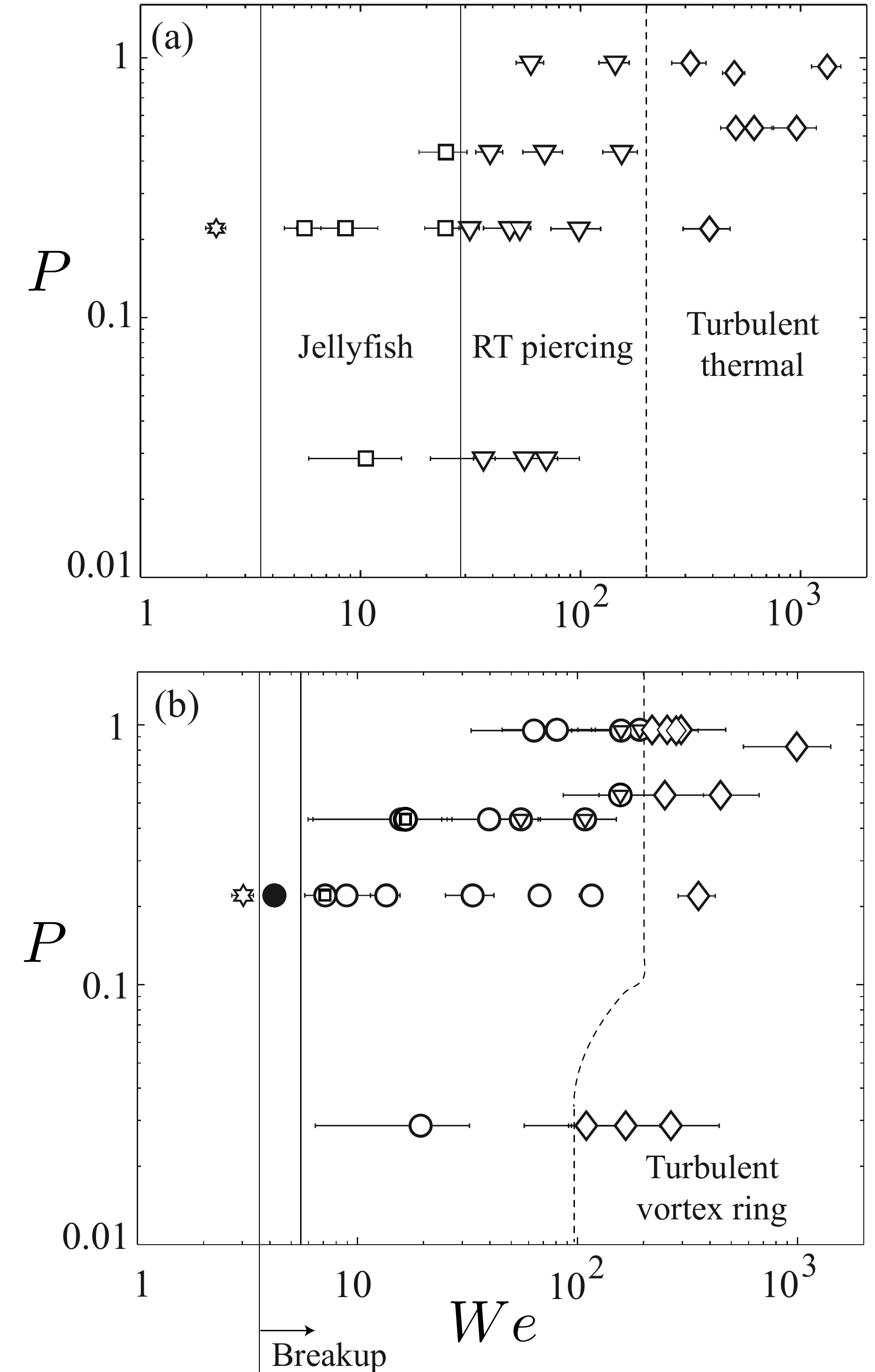}
\end{center}
\caption["Short" caption without tikz code]{\color{black}Fragmentation regimes in $(\Web,\DP)$ space in the Immersed (a) and Surface (b) configurations. Symbols denote: 
\Star[fill=white!1,draw]{0.075}{0.15} ,no fragmentation, oscillating drop; \bicolourcircle{black}{black} , vibrational breakup regime; \bicoloursquare{white}{black}, jellyfish regime; \bicolourtriangle{white}{black} , RT piercing regime;\emptydiamond{black}, turbulent regime; \bicolourcircle{white}{black}, vortex ring destabilization regime;\bicolourtrianglec{black}{white}, intermediate regime between vortex ring destabilization and RT piercing (mushroom-shaped structures, typical of RTI, are observed at the front of a vortex ring);\bicoloursquarec{black}{white}, vortex ring evolving into a jellyfish regime. Plain lines: tentative \color{blue}regime boundaries \color{black}{; dashed lines: progressive transitions.}}
\label{DP_all}
\end{figure}

In \S3.2 it was shown that the initial deformations and their sensitivity to $\DP$ and $\Web$ 
can be qualitatively accounted for by a competition between growth of RTI and advection by the flow. When the latter effect dominates, a vortex ring is formed. 
In the present section the different fragmentation regimes are characterized from the evolution following the initial deformations, and prior to drop formation. The resulting $(\DP,\Web)$ regime diagram, shown in figure \ref{DP_all}, locates the regimes detailed below. 

\subsection{Low and intermediate Weber numbers : wide variety of regimes }

\subsubsection{$We\lesssim6$}

At the lowest Weber numbers ($\Web\approx2-4$) the released fluid takes the form of an oscillating drop. Breakup starts at $\Web\approx5$, \color{blue}consistent with the critical value $We=6$ predicted theoretically by \citet{VillermauxBossa2009}, 
\color{black} and the flow reaches a regime where the released fluid disintegrates into a few large drops as a consequence of large amplitude oscillations at the natural frequency of the drop. This vibrational breakup regime has been documented previously \citep[e.g.][]{PilchandErdman,Gelfand1996}.  

\begin{figure}
\begin{center}
\includegraphics[width=13cm]{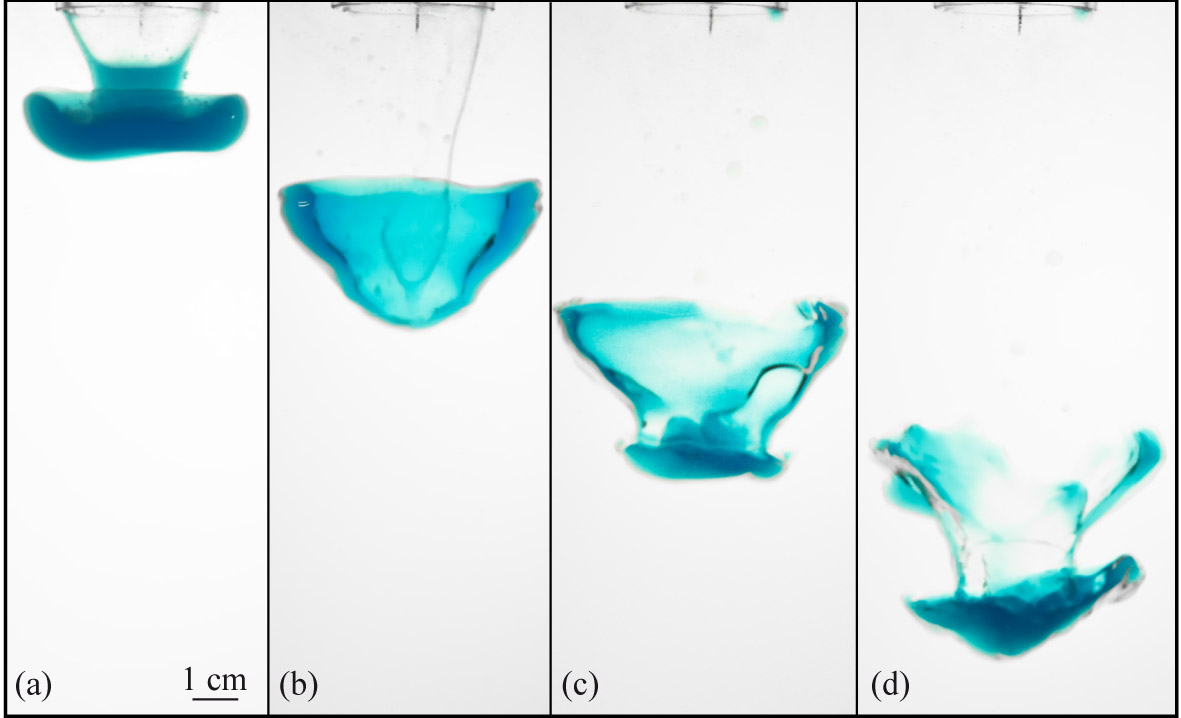}
\end{center}
\caption{\color{black}Experiment in the jellyfish fragmentation regime, $\Web\approx 24$, $\DP\approx 0.22$, Immersed configuration, time intervals of about $0.25$ s.}
\label{Jellyfish}
\end{figure}

%

\begin{figure}
\begin{center}
\includegraphics[width=13cm]{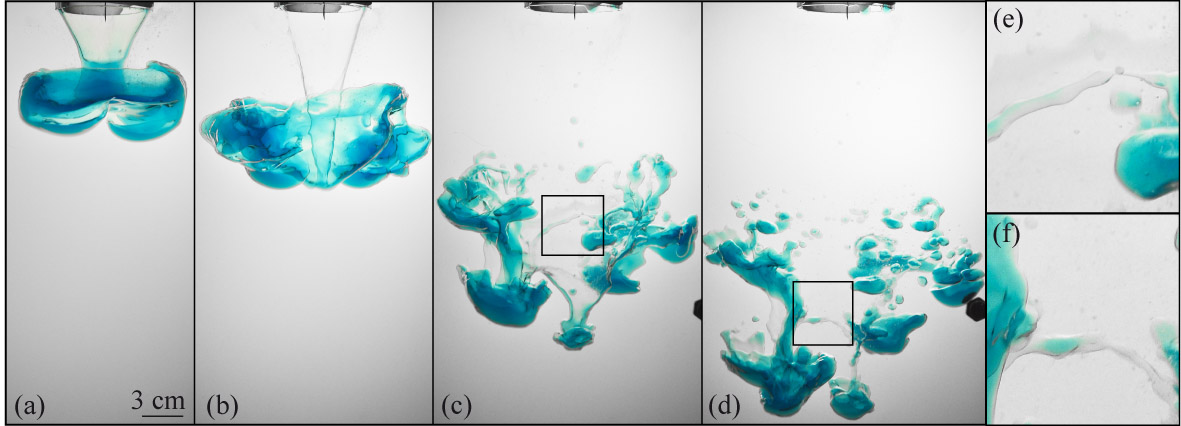}
\end{center}
\caption{\color{black}Experiment in the RT piercing fragmentation regime, $\Web\approx 50$, $\DP\approx 0.22$, Immersed configuration, time intervals of order $0.2$ s. (e,f) Close-ups corresponding to the the square boxes in (c) and (d), respectively.}
\label{RT_frag}
\end{figure}

\begin{figure}
\begin{center}
\includegraphics[width=13cm]{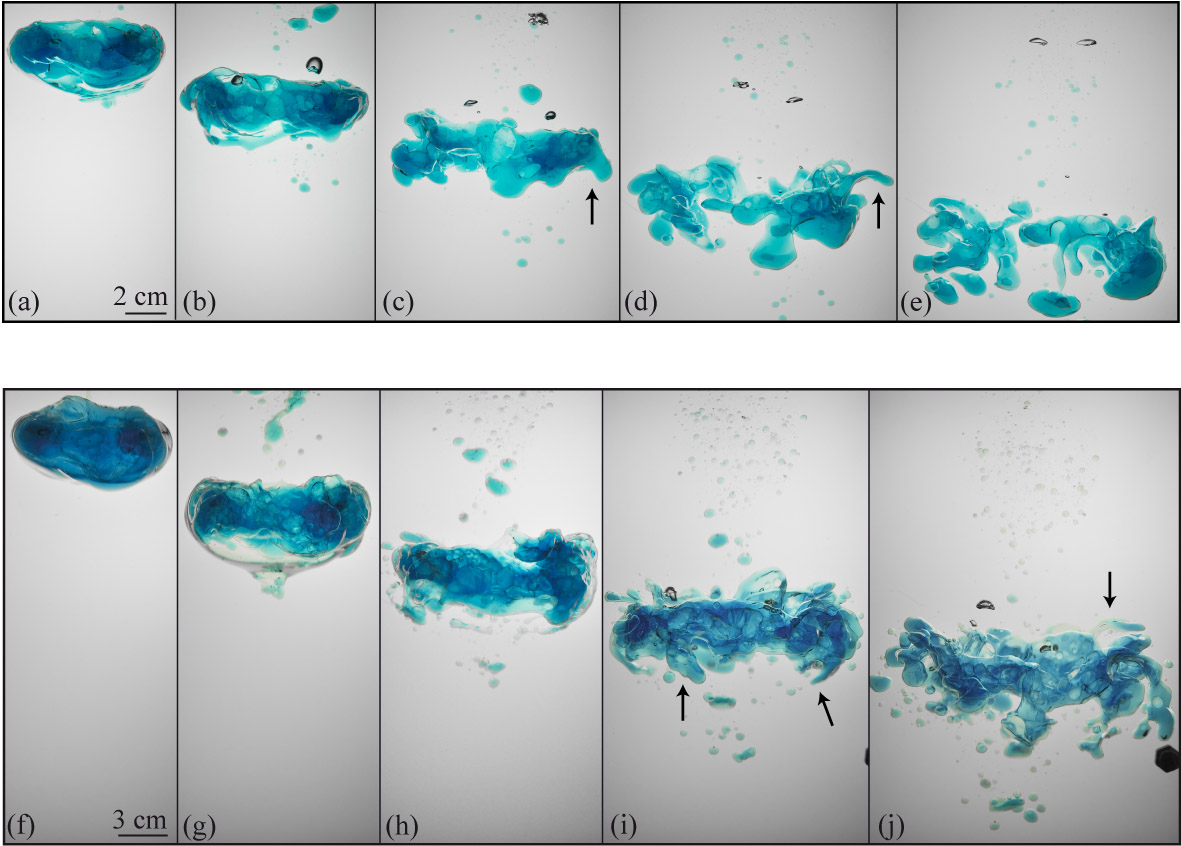}
\end{center}
\caption{\color{black}Experiments in the vortex ring destabilization regime, Surface configuration. (a-e) $\Web\approx 30$, $\DP\approx 0.22$; (f-j) $\Web \approx 70$, $\DP\approx 0.22$. Arrows locate elongated structures or filaments. Time intervals are of about $0.2$ s.}
\label{VR_frag}
\end{figure}

\subsubsection{$We\gtrsim6$; no immiscible ring}
For $\Web\gtrsim 6$, if the released fluid does not roll-up into a ring, 
the evolution that follows the initial deformation and precedes ligament formation is the continuation of the mechanisms identified in \S3.2.     

The fragmentation regime in experiments located below the onset of RTI, 
shown in figure \ref{Jellyfish}, is named the \textit{jellyfish} regime. 
In this regime, the absence of growing RTI allows the flow to remain quasi-axisymmetric until the distance from the tube is equal to a few initial diameters. The initial pancake shape (figure \ref{Jellyfish}(a)) evolves into a U-shaped membrane (figure \ref{Jellyfish}(b)). Then, a portion of released fluid accumulates towards the front, leaving the membrane thinner at the rear (figures \ref{Jellyfish}(c,d)), which leads to the formation of sheared filamentary structures near the rear (figure \ref{Jellyfish}(d)). 
Similar structures, categorized as a shear breakup mode, have been found by \citet{HanTrygg1999} in axisymmetric simulations of drop deformation (see their Fig.5). We note that a vortex ring rolls up in experiments with miscible fluids at similar Reynolds number (in the range $\color{blue}{500-3300}$), suggesting that surface tension prevents the roll-up of the shear layer in these experiments. 
When the initial deformation is dominated by RTI (figure \ref{RT_frag} and Movie 1), the subsequent evolution corresponds to the typical nonlinear evolution of RTI and commonly involves shear instabilities. As a result, the released fluid mass divides into several sub-volumes connected by filamentary structures (figures \ref{RT_frag}(c,d)). The flow shares similarities \color{black} with the multimode breakup regime as described in experiments of aerobreakup and interpreted as a result of RTI \citep{Harper1972,Simpkins1972,Joseph1999,Theofanous2004,Theofanous2008,Zhao2010}. \color{black}Following \citet{Theofanous2004} and \cite{Theofanous2007}, this fragmentation regime is named \textit{RT piercing}. The transition from jellyfish to RT piercing in figure \ref{DP_all}(a) corresponds to the onset of RTI.

\subsubsection{$We\gtrsim6$; with an immiscible ring}
When the initial deformation is dominated by the roll-up of a vortex ring, the evolution prior to ligament formation 
is characterized by the development of additional instabilities on the ring (figure \ref{VR_frag}, Movie 2). 
This \textit{vortex ring destabilization} regime is morphologically different from the RT piercing regime or the jellyfish regime at similar $\Web$ values. \\
A plausible mechanism for the vortex ring destabilization is an elliptical instability, often referred to as the Widnall instability, which has been identified as the mechanism responsible for the destabilization of miscible non-buoyant vortex rings 
\citep{WidnallSullivan1973,Widnall1974,WidnallTsai1977,Saffman1978,Dazin2006}. 
It results from the parametric resonance of neutrally stable modes of vibration, called Kelvin waves, with an underlying quadrupole strain field induced by the vortex ring on itself. 
\cite{HattoriFukumoto2003} and \cite{FukumotoHattori2005} have shown 
that a dipole field resulting from the curvature of the vortex ring can also induce a parametric resonance between two Kelvin waves, called the curvature instability. \cite{Hattori2010} have studied the stability of fat vortex rings, which is the relevant regime for our experiments, where the ratio of the core to vortex ring radius is of order $0.4$. They found that the Widnall instability dominates over the curvature instability, but the combination of the elliptical deformation and the dipole field initiate a third mode of instability whose growth rate exceeds the Widnall instability near the boundary of the ring.\\      
The centrifugal instability is yet another plausible candidate for the destabilization of our immiscible vortex rings. Finally, the presence of a heavy vortex core can also trigger a RT instability where the centrifugal force plays the role of gravity. 
 

The maximum growth rate of the above instabilities are of the same order of magnitude for miscible rings according to previous theoretical and numerical studies \citep{WidnallTsai1977,Hattori2010,Shariff1994,Sipp2005}. Thus, one mechanism can not be 
favored over the others and further investigation, especially accounting for surface tension, 
would be required to identify the dominant mechanisms in our experiments. 
Azimuthal waves are seen in our experiments (e.g. figure \ref{VR_frag}(c)) whereas the most unstable waves of RTI are axisymmetric at small density contrast \citep{Sipp2005}. 

In Surface experiments, transitions from one of the above regimes to another are often progressive. When RTI grows at the front of a developing vortex ring, the flow is a combination of RT piercing and vortex ring destabilization regimes (figure \ref{DP_all}) and, in a few and isolated experiments at $\Web\leq 20$, a vortex ring forms but finally evolves into a jellyfish fragmentation regime (figure \ref{DP_all}).

\color{blue} We note that air entrained during the formation of some immiscible vortex rings is responsible for the rising air bubbles seen in figure \ref{VR_frag}. Comparison between experiments conducted at the same parameter values, with and without air bubbles, 
indicates that the fragmentation regime is little affected by the presence of air bubbles. \color{black}

\subsection{High Weber numbers: turbulent regime}
%
%
%

\begin{figure}
\begin{center}
\includegraphics[width=12.1cm]{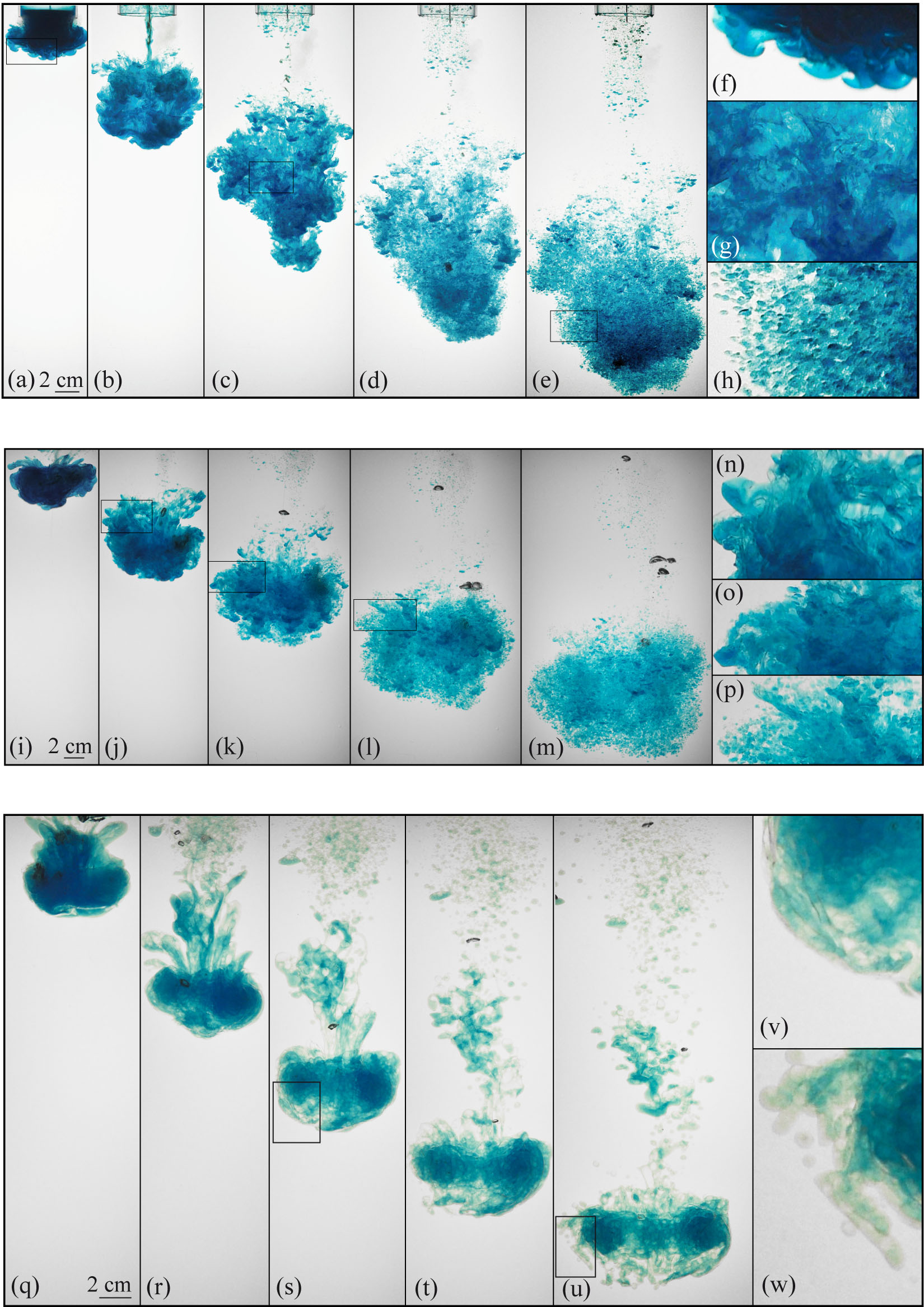}
\end{center}
\caption{Turbulent regime. (a-h) Immiscible turbulent thermal, $\Web\approx 10^3$, $\DP\approx 0.92$, Immersed configuration, time intervals of about $0.2$ s. (i-w) Immiscible turbulent buoyant vortex rings, Surface configuration. (i-p) $\Web\approx 10^3$, $\DP\approx 0.82$, time intervals of about $0.2$ s; (q-w) $\Web\approx 200$, $\DP\approx 0.03$, time intervals of about $0.4$ s. (f,g,h) (n,o,p) (v,w) Close-ups corresponding to the square boxes in (a,c,e), (j,k,l) and (s,u), respectively.}
\label{Turb}
\end{figure}

When $\Web$ is increased above $\sim 100$, a progressive transition leads to the turbulent regimes illustrated in figure \ref{Turb} (see also Movie 3). 
The deformations of the immiscible interface are chaotic and exhibit a wide range of length scales (e.g. figures \ref{Turb}(c,g)). In the experiment shown in figures \ref{Turb}(a-h) 
initial deformations are dominated by RTI (seen in (a) and (f)) whereas no RTI develops in the experiment shown in figures \ref{Turb}(q-w) 
(Surface configuration, low $\DP$). The initial deformations in figures \ref{Turb}(i-p) 
are more ambiguous: the waves in (i) do not have a clear mushroom-shaped structure as in figures \ref{Turb}(a,f), 
but their characteristic size is consistent with the predicted wavelength for RTI and this experiment is located in a region of parameter space where we expect RTI to emerge according to results from \S3.2. 
Despite the different initial deformations, the large-scale flow has common features in the three experiments: the released fluid is contained inside a coherent structure whose shape is self-similar during the fall and which grows by entrainment of ambient fluid. 
This behavior is similar to the case of a fluid mass evolving in another miscible fluid at high Reynolds number, as described by \citet{Batchelor1954} and \citet{Scorer1957} for thermals, \citet{Maxworthy1974} and \citet{GlezerColes1990} for non-buoyant vortex rings, and \cite{Turner1957} for buoyant vortex rings. \color{blue}{Again, the fragmentation regime is little affected by the presence of air bubbles in the Surface configuration (figures \ref{Turb}(i-w)).} \color{black}

The geometry of the coherent structure in figures \ref{Turb}(q-w) 
can be approximated by an oblate spheroid of large width to height ratio ($\approx1.8$), much like miscible non-buoyant vortex rings. In contrast, the coherent structure in figures \ref{Turb}(a-h) can be approximated by a prolate spheroid much like the shape of miscible turbulent thermals.

\begin{figure}
\begin{center}
\includegraphics[width=6.5cm]{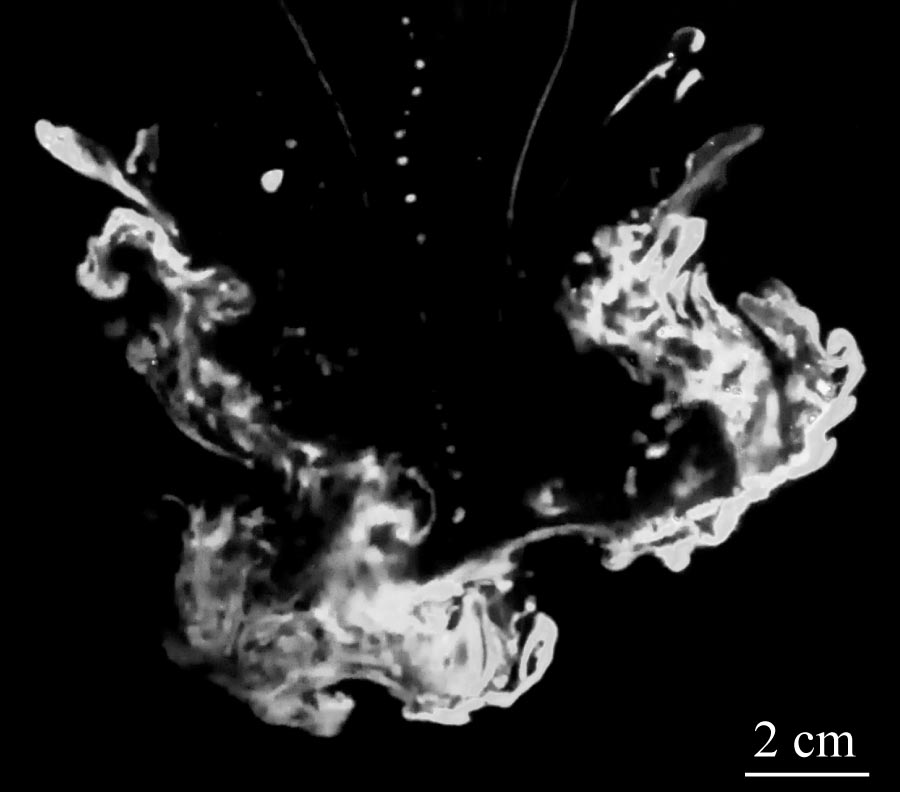}
\end{center}
\caption{
Cross-section of an immiscible turbulent thermal, obtained using light-induced fluorescence, modified from \citet{Deguen2013}. $\Web\approx 10^3$, $\DP\approx 0.54$, Immersed, $z\approx R$.} 
\label{LS}
\end{figure}

A cross-section of an immiscible thermal is shown in figure \ref{LS}. 
It reveals small-scale intermingling between released and ambient fluids in the entire thermal, even though the two immiscible phases remain continuous. This demonstrates that ambient fluid is entrained in the thermal before the released fluid breaks into fragments. The immiscible interface has a fractal structure as demonstrated in \citet{Deguen2013}. Comparison between figure \ref{LS} and images obtained in equivalent miscible experiments \citep[see their figure 4, left, which shows a turbulent thermal with $Re\approx 5000$, $P\approx0.05$, at $z\approx3R$]{BondJohari2010} demonstrates 
that the large-scale internal structure of turbulent thermals is morphologically similar in miscible and immiscible experiments. 




\section{Final fragmentation stage: breakup}

\subsection{Description of the physical processes}
%
%
%
%
%
%
%

As in other fluid fragmentation processes \citep{Hinze1955}, 
the deformations identified \color{blue}in \color{black} \S3 and \S4 result in the formation of elongated and filamentary structures, or liquid ligaments, (e.g. figure \ref{Jellyfish}(d), \ref{RT_frag}(c,d,e,f), \ref{VR_frag}(d), \ref{Turb}(w)) 
and their destabilization, probably through capillary instabilities, leads to breakup. However 
the spatial distribution and formation time of these ligaments differ from one fragmentation regime to the other. 

In the jellyfish (figure \ref{Jellyfish}(d)) or the RT piercing (figure \ref{RT_frag}(c,d,e,f) and Movie 1) regimes thin filamentary structures connect larger blobs of released fluid. 
In the vortex ring destabilization regime the azimuthal waves result in the formation of thinner portions on the ring 
(figure \ref{VR_frag}(c)), which eventually break the ring in separated blobs (between figures \ref{VR_frag}(c) and \ref{VR_frag}(d)). In the meantime, the azimuthal disturbances are stretched by the mean shear flow leading to the formation of spiraling filaments located preferentially on the ring boundaries (arrows in figure \ref{VR_frag}, Movie 2). 

In the turbulent vortex ring regime at $P\approx 0.03$, ligaments form at the external boundary of the ring (figure \ref{Turb}(u,w)). The ligament formation, followed by their breakup, is a multi-step process: the ring is progressively peeled, whereas the 
primary breakup of the entire released fluid volume occurs in a single and brief event in the turbulent regime for $P \gtrsim 0.2$ (between figures \ref{Turb}(d) and \ref{Turb}(e), and between figures \ref{Turb}(j) and \ref{Turb}(l), see also Movie 3). 
In this case, as can be inferred from figures \ref{Turb}(b,c) and from the cross-section in figure \ref{LS}, breakup probably results from capillary instabilities on filamentary structures stretched by the turbulent flow in the entire thermal volume. However, higher temporal and spatial resolution is required to test this interpretation. 

\subsection{Breakup length}

%

\begin{figure}
\begin{center}
\includegraphics[width=13.5cm]{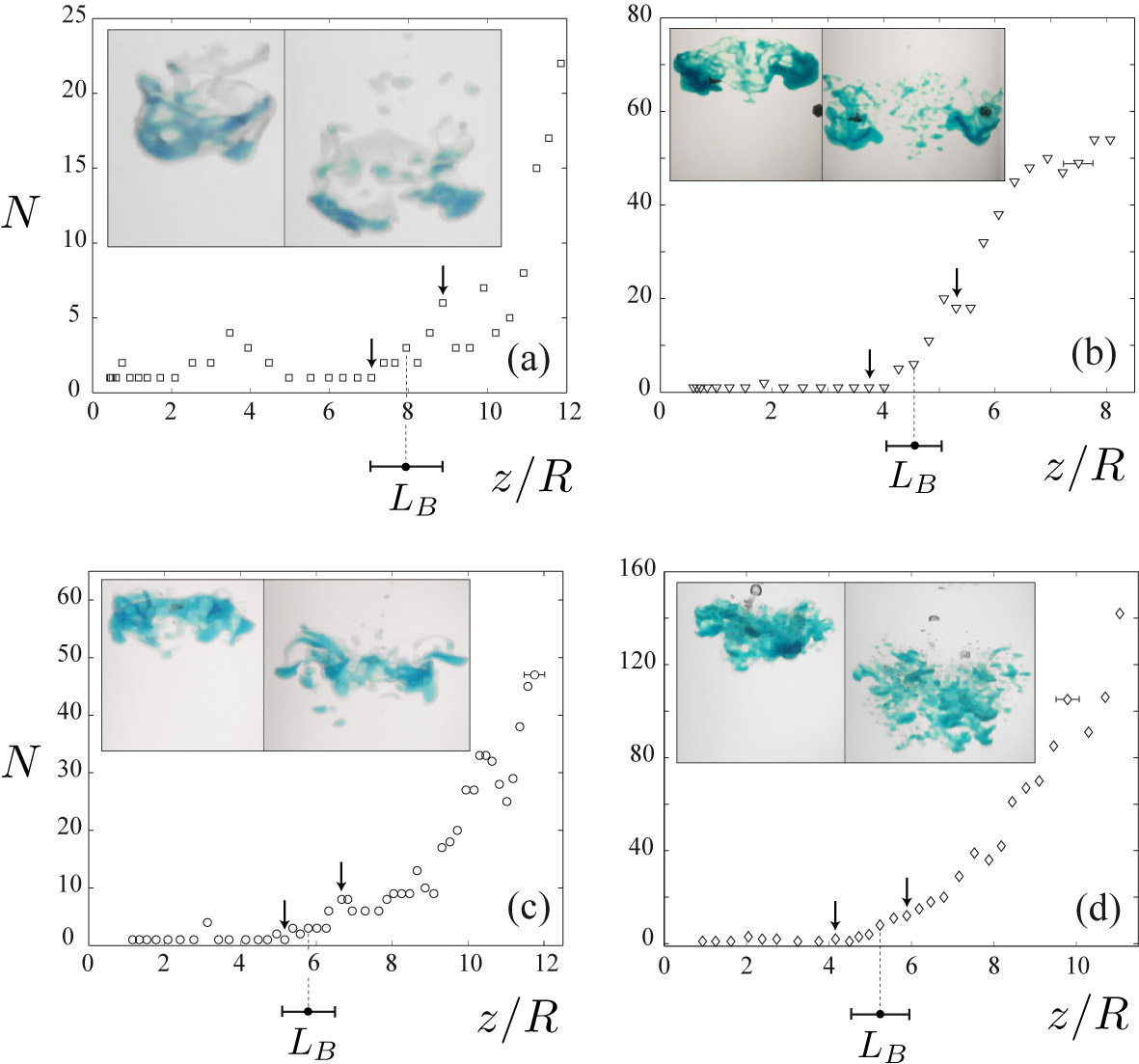}
\end{center}
\caption{\color{blue}{Number of connected objects $N$ in binary images as a function of dimensionless distance from the tube in four single experiments. Symbols shapes refer to the fragmentation regime (as in figure \ref{DP_all}). The breakup length $L_B$ is the dimensionless distance from the tube at which $N$ starts to increase. Error bars indicate typical measurement uncertainties on $L_B$ and $z/R$. Error bars on $L_B$ are large enough such that they incorporate spurious variations in $N$ due to hidden droplets or air bubbles. The inserts show backlighting images of the released fluid before and after $L_B$, at distances located by the arrows. (a) Jellyfish regime, $\Web\approx 29$, $\DP\approx0.22$, Immersed configuration. (b) RT piercing regime, $\Web\approx 170$, $\DP\approx0.43$, Immersed configuration. (c) Vortex ring destabilization regime, $\Web\approx 45$, $\DP\approx0.22$, Surface configuration. (d) Turbulent regime, $\Web\approx 280$, $\DP\approx0.96$, Surface configuration.}\color{black}} \color{black}
\label{LB_def}
\end{figure}

The dimensionless breakup length $L_B$, is defined as the dimensionless distance from the tube at which the number of connected objects in binary images starts to increase \color{blue} (see figure \ref{LB_def})\color{black}. It marks the beginning of primary breakup.
\color{blue}{Most of the drops formed in the rear of the released fluid from the rupture of a membrane that remains attached to the tube (e.g. figures \ref{Jellyfish}(b,c), figure \ref{RT_frag}(b)) or from breakup in the wake of turbulent thermals or vortex rings (e.g. figure \ref{Turb}) are automatically excluded from binary images; a few remain and cause local, spurious variations in the number of connected objects, as seen close to $z/R\approx 3$ in figures \ref{LB_def}(a) and \ref{LB_def}(c). These are not taken into account in the determination of $L_B$. \color{blue}{The sporadic presence of air bubbles in Surface experiments retards the released fluid at early times, and may thus decrease the breakup length, while increasing its standard deviation. 
}}\color{black}

\begin{figure}
\begin{center}
\includegraphics[width=13.5cm]{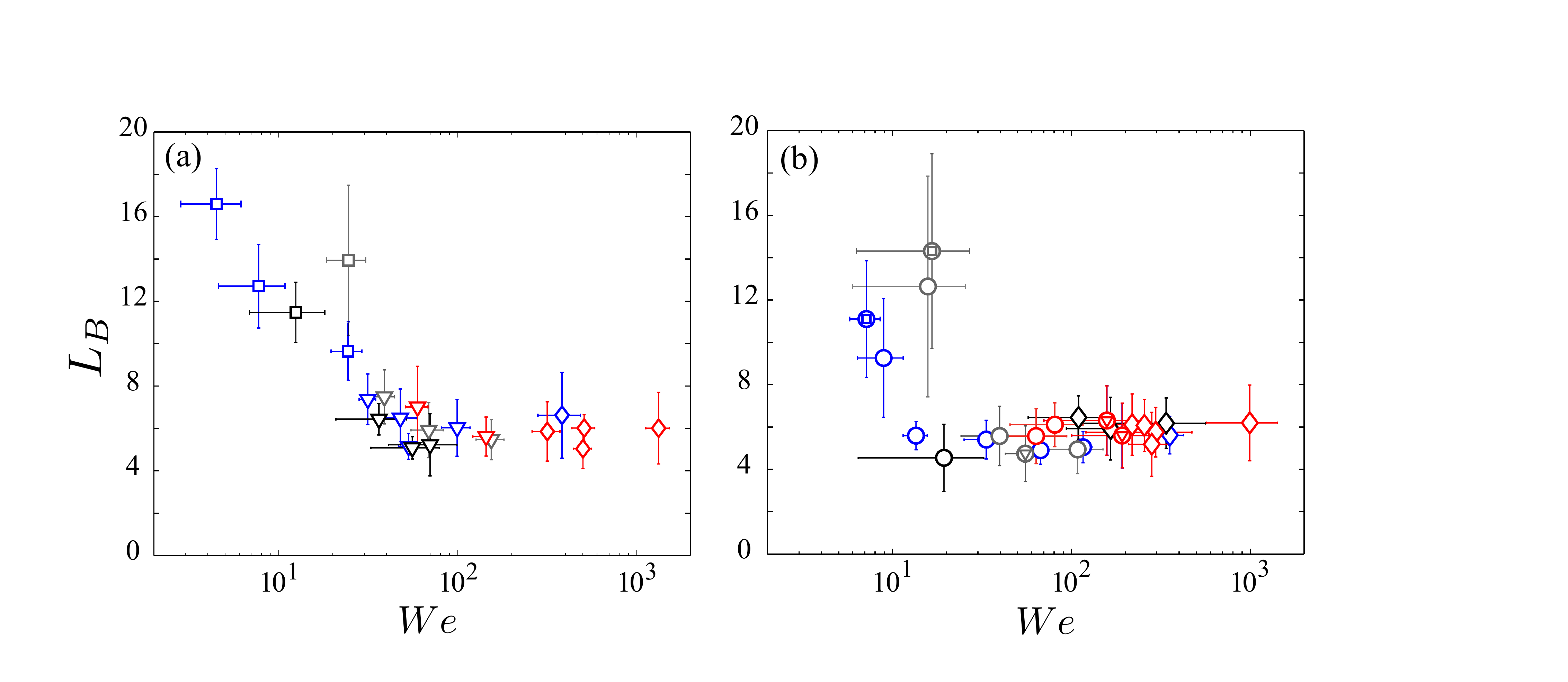}
\end{center}
\caption{Dimensionless breakup length as a function of Weber number in (a) Immersed and (b) Surface experiments. Symbol shapes: as in figure \ref{DP_all}. Black: $P\approx0.03$; blue: $P\approx0.22$; grey: $0.43\leq P \leq 0.54 $; red: $ 0.83 \leq P \leq 0.96 $. 
Insert in (a): number of connected objects $N$ in binary images as a function of dimensionless distance from the tube in a given experiment.}
\label{LB}
\end{figure}

In Immersed experiments, the jellyfish, RT piercing and turbulent regimes correspond \color{black}to specific regions in figure \ref{LB}(a). For a given $P$ value in the jellyfish and RT piercing regimes, the overall trend of the dimensionless breakup length is a decrease with increasing $We$. In the turbulent regime, the variation of $L_B$ with $We$ is within the experimental error and in the range $4.5-7.5$. 

In Surface experiments (figure \ref{LB}(b)), the different regimes overlap, with no distinctive behavior from one regime to the other, suggesting common destabilizing mechanisms. 
Given experimental errors, no significant variation of $L_B$ is seen for $\color{blue}{We\gtrsim 25}$: the different fragmentation regimes collapse between $L_B \approx 4.5$ and $L_B\approx 7$.

\newpage
\section{\color{black}{Integral model for the turbulent regime}}



\color{black}It has been shown in \S4 that the flow takes the form of turbulent vortex rings (Surface experiments) and turbulent thermals (Immersed experiments) for $We\gtrsim 200$. 
Following \citet{Deguen2011}, we assume that immiscibility does not affect the macroscopic behavior of such structures, \color{blue}{so that we can apply models that have been developed in the context of miscible fluids \citep{Morton1956,Maxworthy1974,Escudier1973,Thompson2000} and particle clouds \citep{Bush2003} at high Reynolds numbers}\color{black}. 
In the present section we consider the general case of buoyant vortex rings, allowing for initial momentum and large density differences between the ring and the ambient fluid. $Re$ is in the range $\color{blue}{3000}\color{black}-11000$ in the experiments considered in this section.

\subsection{Theoretical considerations}



Following the turbulent entrainment hypothesis \citep{Taylor1945,Morton1956}, we assume that the rate of growth of the vortex ring mass is proportional to its velocity and its surface area:   

\begin{eqnarray}
\dfrac{d}{dt} \left \lbrack  \dfrac{4}{3}\pi \rho r^3 c_1 \right\rbrack   = 4 \pi r^2   c_2  \alpha \rho_a u,
\label{Mass}
\end{eqnarray}
where $\rho$ is the mean density of the ring, $u$ is the ring velocity, $c_1$ and $c_2$ are shape factors which relate the actual volume $V$ and surface area of the ring to those of an equivalent sphere of radius $r$, $r$ being a measure of the size of the moving fluid mass, and $\alpha$ the entrainment coefficient as introduced in \citet{Taylor1945} and \citet{Morton1956}.\\ 
Making use of the relations $\rho r^3= \Delta \rho R^3/c_1+\rho_a r^3$ and $u = d z/dt$, the mass conservation equation (\ref{Mass}), in the absence of density stratification, becomes

\begin{eqnarray}
\color{black}
\frac{d r}{d z} = \alpha^\prime,
\label{Rz}
\end{eqnarray}
\color{black}
where $\alpha^\prime = \alpha c_2/c_1$. Equation (\ref{Rz}) implies that $r\propto z$ at all times whenever the entrainment coefficient $\alpha^\prime$ is constant. This linear relationship between $r$ and $z$ can be derived from dimensional analysis in the special case of a Boussinesq thermal \citep{Batchelor1954} or a non-buoyant vortex ring \citep{Maxworthy1974}.



In the absence of density stratification the total buoyancy $b = (\rho - \rho_a)/\rho_a g V$ of the moving fluid mass is conserved and equal to its initial value $B$. Then, the impulse conservation equation takes the form 

\begin{eqnarray}
\dfrac{d}{dt} \left \lbrack\frac{4}{3}\pi (\rho + k \rho_a)  r^3 c_1 u  \right\rbrack  = \rho_a B - \frac{1}{2} C_D^\prime \rho_a u^2\pi r^2,
\label{Momentum}
\end{eqnarray}
where $C_D^\prime = C_D c_3$, $c_3$ is another shape factor, $C_D$ is the drag coefficient and the added mass coefficient $k$ accounts for the change 
in kinetic energy of the surrounding fluid \citep{SaffmanBook,Escudier1973}.


Using the equivalent radius of the released fluid $R$ as a length scale and $R^2\sqrt{(4 \pi/3) /B}$ as a time scale, the final set of non-dimensional equations takes the form


\begin{eqnarray}
\left \lbrack \DP + (1+k) c_1 \tilde{r}^3 \right \rbrack \dfrac{d \tilde{u}}{d\tilde{t}}  = 1 - 3 \alpha^\prime  \left \lbrack c_1(1+k)+\dfrac{C_D^\prime}{8\alpha^\prime} \right \rbrack \tilde{r}^2 \tilde{u}^2
\label{Momentum2},\\
\dfrac{d \tilde{r}}{d \tilde{t}} = \alpha^\prime \dfrac{d \tilde{z}}{d \tilde{t}}, \hspace{1cm}\dfrac{d \tilde{z}}{d \tilde{t}} =  \tilde{u}.
\label{Rz2}
\end{eqnarray}

Equation (\ref{Momentum2})-(\ref{Rz2}) can be integrated in time if $\alpha^\prime$, $C_D^\prime$, $k$, $c_1$ and the initial conditions on $\tilde{u}$, $\tilde{r}$ and $\tilde{z}$ are given.


Since $d\tilde{r} /dt = \alpha^\prime \tilde{u}$, (\ref{Momentum2}) becomes

\begin{eqnarray}
\dfrac{\alpha^\prime}{2}\left \lbrack \DP+ (1+k) c_1 \tilde{r}^3 \right \rbrack  \dfrac{d \tilde{u}^2}{d\tilde{r}}  = 1 - 3 \alpha  \left \lbrack c_1(1+k)+\dfrac{C_D^\prime}{8\alpha} \right \rbrack \tilde{r}^2 \tilde{u}^2.
\label{Momentum3}
\end{eqnarray}
For constant values of $C_D^\prime$, $k$, $c_1$ and $\alpha^\prime$, the general solution of the first-order linear differential equation (\ref{Momentum3}) is
\begin{eqnarray}
\tilde{u}^2 = \dfrac{2}{\alpha^\prime} \intevariable{\tilde{r}_0}{\tilde{r}} {\dfrac{\left (\DP + (1+k) c_1 x^3 \right )^{\gamma-1}}{\left (\DP + (1+k) c_1 \tilde{r}^3 \right )^{\gamma}} \mathrm{d} x} + {\tilde{u}_0}^2 \left \lbrack \dfrac{\DP + (1+k) c_1 \tilde{r_0}^3 }{P+ (1+k) c_1 \tilde{r}^3 }\right \rbrack ^{\gamma},
\label{GeneralSolution}
\end{eqnarray}
where $\gamma = 2+ C_D^\prime/\left(4\alpha^\prime(1+k)c_1 \right )$ and the subscript $_0$ denotes initial conditions. 
Closed-form solutions for $\tilde{u}$ exist if $C_D^\prime=0$ or, if the Boussinesq approximation is valid ($P \ll 1$), for arbitrary values of $C_D^\prime$ (given in appendix \ref{TheoreticalSol}). \\
In the following limit: 
\begin{eqnarray}
 \tilde{z}-\tilde{z}_0\gg \frac{\tilde{r}_0}{\alpha^\prime} \text{  and  } \tilde{z}-\tilde{z}_0\gg \frac{\tilde{r}_c}{\alpha^\prime} \text{  where  } \tilde{r}_c \text{  satisfies  } (1+k)c_1{\tilde{r}_c}^3\gg \DP,
 \nonumber
 \label{Cond1}
\end{eqnarray}
 the solution (\ref{GeneralSolution}) has an asymptote given by 
\begin{eqnarray}
\tilde{u}^2\approx\dfrac{2}{{\alpha^\prime}^3 (1+k) c_1(3\gamma -2)} \dfrac{1}{\left(\tilde{z} - \tilde{z}_0\right)^2}\left \lbrack 1 + \left (\dfrac{L_M}{\tilde{z}-\tilde{z}_0} \right ) ^{3\gamma-2}\right \rbrack,
 \label{Approx2}
\end{eqnarray}
where $L_M$ is given by 
\begin{eqnarray}
L_M = \left (\dfrac{1}{2}{\alpha^\prime }^3 (1+k)c_1(3\gamma - 2) {\tilde{u}_0}^2\right )^{1/3\gamma-2} \left (  \dfrac{\DP + (1+k) c_1 \tilde{r_0}^3 }{(1+k) c_1 {\alpha^\prime}^3}\right )^{\gamma/3\gamma-2}.
\label{LM}
\end{eqnarray}
Since \hbox{$3\gamma - 2 > 0$}, $L_M$ is the distance over which the initial momentum affects the solution, often called the Morton length. 
If $\tilde{z}-\tilde{z}_0 \gg L_M$ the initial momentum becomes inconsequential and the flow reaches the same asymptotic regime as in thermals, i.e. in terms of dimensional variables

\begin{eqnarray}
\label{Asymptotic1}
u(z) &\approx& f\frac{\sqrt{B}}{z-z_0},\\
\label{Asymptotic2}
\dfrac{d\left (z-z_0\right ) ^2}{dt} &\approx& 2 f \sqrt{B}, \\
\label{Asymptotic3}
\text{       where  } f &=&\left\{8/3\pi c_1 (1+k) \alpha^3 + C_D^\prime/2 \pi \alpha^2 \right\}^{-1/2}. 
\end{eqnarray}

%


In miscible turbulent thermals \citep{Scorer1957,Richards1961,Thompson2000} or in non-buoyant vortex rings \citep{Maxworthy1974,GlezerColes1990} the size of the structure grows linearly with depth as predicted by (\ref{Rz}) for a constant $\alpha^\prime$ value. In miscible thermals the entrainment coefficient is usually determined by measuring the growth of the thermal half-width and typically $\alpha_T=0.25\pm0.1$, where $\alpha_T$ is the entrainment coefficient for thermals. The entrainment coefficient of non-buoyant vortex rings, $\alpha_V$, is commonly determined by measuring the growth of the radius of the vortex ring core and it can be described as $\alpha_V=0.01\pm0.005$. 
The entrainment coefficients of buoyant vortex rings were not directly reported by \citet{Turner1957} but values ranging from $0.02$ to $0.18$ can be extracted from his figure 3 and other parameter estimations. These values lie between $\alpha_V$ and $\alpha_T$, the lowest values being reached when the ratio of initial impulse to buoyancy force is the highest, i.e. when the initial momentum dominates the total momentum. 
\color{blue}{Given these observations, the entrainment coefficient must vary with time in a buoyant vortex ring since the flow eventually behaves as a thermal as predicted by (\ref{Asymptotic1})-(\ref{Asymptotic3}) and $\alpha^\prime$ is equal to $\alpha_T$ in this asymptotic regime. }\color{black} 
From a theoretical point of view, it is important to \color{blue}account for these effects in a self-consistent model of buoyant vortex rings\color{black}.

\color{blue}{An important difference between non-buoyant vortex rings and thermals is the presence of baroclinic vorticity generation in the latter case, which affects the vorticity distribution and thus the entrainment coefficient \citep{Lundgren1992,Alahyari1995,Alahyari1997,BondJohari2010}. In positively buoyant vortex rings, the amplitude of baroclinically-generated vorticity is positively correlated with the local Richardson number, representing the ratio of buoyancy to inertial forces, and evolving from values much smaller than $1$ 
near the source, where the excess in initial momentum dominates the total momentum, to values of order $1$ in the asymptotic thermal-like regime, where the initial momentum has become inconsequential with respect to the buoyancy-induced momentum. 
During this transition the entrainment coefficient should vary from $\alpha_V$ to $\alpha_T$. 
Using the following definition of the Richardson number
\begin{eqnarray}
Ri = \dfrac{\Delta \rho g R^3}{\rho u^2 r^2},
\label{Ri}
\end{eqnarray}
$Ri$ varies from $0$ in non-buoyant vortex rings to a constant value \hbox{$Ri_T=2c_1(1+k)\alpha_T+\frac{3}{8}C_D^\prime$} in Boussinesq thermals when $r\gg R$ (the asymptotic regime given by equations (\ref{Asymptotic1})-(\ref{Asymptotic3})). 

\citet{Turner1957} showed that the entrainment coefficient of a buoyant vortex ring in which the circulation $K$ remains constant is proportional to $B/K^2$, the ratio of buoyancy to inertial forces, i.e. a Richardson number. 
The circulation of a buoyant vortex ring is probably not constant; observations in non-buoyant vortex rings indicate it is lost to the wake by shedding of vortical structures 
\citep{Maxworthy1972,Maxworthy1974,GlezerColes1990,WeigandGharib1994,Dazin2006NL,Archer2008}. A significant wake is observed in immiscible buoyant vortex rings at low Richardson numbers (figures \ref{Turb}(q-w), $Ri$ down to $0.1-0.2$) whereas almost no wake is present at Richardson number of order $1$ in immiscible thermals (figures \ref{Turb}(a-h)), in agreement with previous studies on 
miscible non-buoyant vortex rings and thermals \citep{Maxworthy1972,Maxworthy1974,Scorer1957,BondJohari2010}. The assumption of constant circulation also rests on vorticity being confined to a region that does not extend to the centre of the ring, such that 
the contribution of baroclinic vorticity generation to the total ring circulation is equal to zero. 
This contribution is in general different from zero in the case of thick-core buoyant vortex rings. 
Despite the above limitations, Turner's result gives a physical argument in favour of a linear relationship between $\alpha^\prime$ and $Ri$. 

Such a linear relationship is also expected from the analogy with turbulent buoyant jets. In turbulent buoyant jets the entrainment coefficient varies during the transition from a jet-like to a plume-like behavior \citep{Fisher1979,WangLaw2002}. 
Theoretical parameterizations, one inspired by the work of \citet{PriestleyBall1955} and the other developed by \citet{Kaminski2005}, predict that the entrainment coefficient is a linear function of a local Richardson number. }
\color{black}
Accordingly, a natural parameterization to account for variations of $\alpha^\prime$ in buoyant vortex rings is 
\begin{eqnarray}
\alpha^\prime = \alpha_V + \left(\alpha_T-\alpha_V \right )\dfrac{Ri}{Ri_T}.
\label{AlphaRiP}
\end{eqnarray}

Equations (\ref{Momentum2})-(\ref{Rz2}) remain unchanged if $\alpha^\prime$ varies with time. Thus, (\ref{Momentum2})-(\ref{Rz2}) and (\ref{AlphaRiP}) can be coupled and integrated forward in time, giving a self-consistent model for the evolution of a buoyant vortex ring. It is important to emphasize that, once the parameterization between $\alpha^\prime$ and $Ri$ is specified, the above model has one free parameter \color{black}less than in the case of a constant entrainment coefficient. For instance, in the case of parameterization (\ref{AlphaRiP}), local values of $Ri$ and $\alpha^\prime$ can be experimentally determined, which leads to an estimation of $Ri_T$. From the estimation of $Ri_T$ we obtain a linear relationship between $C_D^\prime$ and $k$, whereas these parameters are independent in the case of a constant entrainment coefficient.

\color{black}
\subsection{\color{black}Experimental results - comparison with theory}

\color{black}{I\color{black}n this section, the analogy with miscible turbulent thermals and vortex rings is tested by comparing results from immiscible fluid experiments with both theoretical predictions and experimental results obtained with miscible fluids. }\\ 


\color{black}
\subsubsection{Entrainment coefficient}\label{Coef}

\begin{figure}
\begin{center}
\includegraphics[width=9cm]{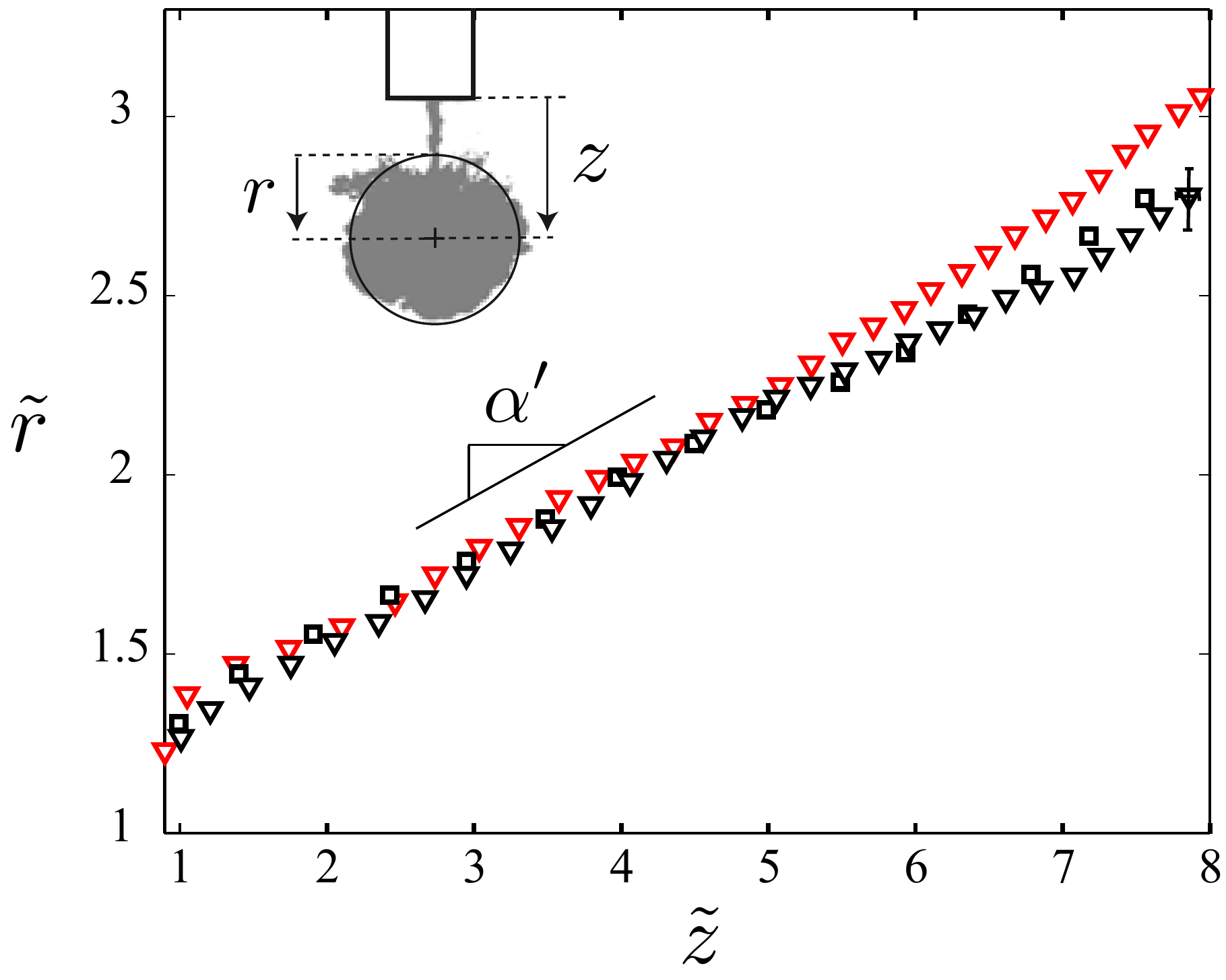}
\end{center}
\caption["Short" caption without tikz code]{\color{black}Dimensionless equivalent radius as a function of the dimensionless distance between the centroid and the tube in Immersed experiments;  \emptytriangle{black} , $\DP\approx0.92$; \emptysquare{black} , $\DP\approx0.22$; \emptytriangle{red} , miscible fluids, $\DP \approx 0.19$. \color{blue}{Characteristic error bars associated with measurement uncertainties are shown}. \color{black}}
\label{AlphaEx}
\end{figure}

Vortex ring equivalent radius and centroid are estimated from video images as described in \S2, considering the largest connected object in the image for the equivalent radius. 
In our immiscible thermals (Immersed experiments), the equivalent radius evolves linearly with the distance traveled, in agreement with equation (\ref{Rz}) and with the turbulent entrainment hypothesis, as illustrated in figure \ref{AlphaEx}. As shown in this figure, experiments with miscible and immiscible fluids have very similar behaviors, supporting the analogy with miscible thermals.

For each experiment an entrainment coefficient $\alpha^\prime$ is estimated. As pointed out in previous studies \citep{Scorer1957,Richards1961,Thompson2000,Bush2003} a large variability in $\alpha^\prime$ between successive realizations 
is unavoidable and inherent to this turbulent flow, which is not quasi-stationary in the reference frame of the laboratory. 
The mean values of $\alpha^\prime$ in Immersed experiments are reported in table \ref{alpha}. Uncertainties take into account both the uncertainty on $\alpha^\prime$ in each experiment and the variability between experiments. Note that the measured entrainment coefficient is $\alpha^\prime = \alpha c_2/c_1$, which depends in principle on the method used to measure the radius and the position of the thermal through the coefficients $c_1$ and $c_2$. 
In our miscible fluid experiments we find $\alpha^\prime = 0.24 \pm 0.05$ (table \ref{alpha}), in agreement with previously published studies in which the maximal half-width of the thermal (rather than the equivalent radius) is used to estimate $r$. 
The use of the equivalent radius is favored in this study because the resulting signal is much smoother than when using the maximal width, which is very sensitive to local deviations from the self-similar behavior. 
In our immiscible thermals $\alpha^\prime$ is slightly lower at $\DP\approx 0.22$ but, given the uncertainties, no significant variations of the entrainment coefficients with the \color{blue}{normalised density difference $\DP$ }\color{black} is observed (table \ref{alpha}). We conclude \color{black} that the entrainment coefficient in our immiscible thermals, $\alpha^\prime=\alpha_T$, is such that $\alpha_T = 0.23\pm0.06$, with no significant deviation from miscible thermals.

\color{black}

%

\begin{table}
\begin{center}
\small
\begin{tabular}{rrr}
{} & {$\DP$}&{$\alpha^\prime$}\\
{} & {}&{}\\
{\textbf{Immiscible}} & {$0.220\pm0.001$}&{$0.20\pm0.03$}\\
{} & {$0.536\pm0.002$}&{$0.25\pm0.03$}\\
{} & {$0.954\pm0.002$}&{$0.24\pm0.05$}\\
{\textbf{Miscible}} & {$0.192\pm0.001$}&{$0.24\pm0.05$}\\
\end{tabular}
\caption{Values of the entrainment coefficient $\alpha^\prime$ in Immersed experiments at different \color{blue}{normalised density difference $\DP$}\color{black}, with miscible and immiscible fluids. Errors represent standard deviations in series of experiments.
}
\label{alpha} 
\end{center}
\end{table}

\begin{figure}
\begin{center}
\includegraphics[width=10cm]{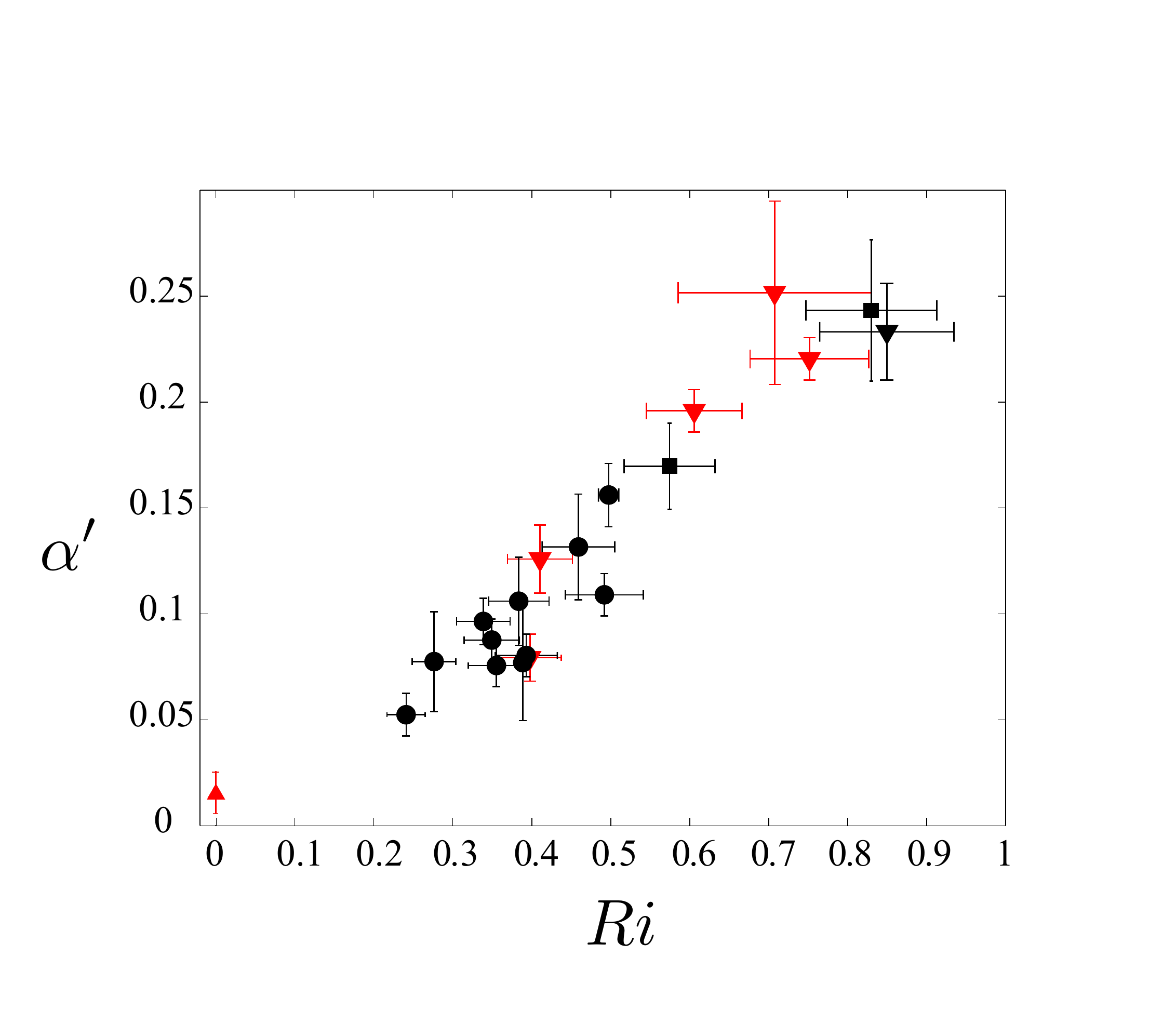}
\end{center}
\caption["Short" caption without tikz code]{Measured turbulent entrainment coefficient as a function of the local Richardson number in Surface experiments. \coloursquare{black} , $\DP \approx 0.82$; \colourtriangle{black} , $\DP\approx0.22$; \colourcircle{black} , $\DP\approx 0.03$; \colourtriangle{red} , miscible fluids, $\DP \approx 0.19$; \colourtriangleup{red} , miscible fluids, $\DP = 0$. \color{blue}{Error bars indicate typical measurement uncertainties.} \color{black}
}
\label{AlphaRi}
\end{figure}



In Surface experiments, the equivalent radius varies linearly with depth, at least locally, \color{black}{and local values of $\alpha^\prime$ can be estimated. 
Figure \ref{AlphaRi} shows that a wide range of $\alpha^\prime$ values are found (from $\sim 0.05$ to $\sim 0.25$). 
It also illustrates that the local Richardson number is a control parameter of the entrainment coefficient. By definition, $\alpha^\prime=\alpha_V$ in miscible fluid experiments with no initial buoyancy ($Ri=0$, bottom-left corner in figure \color{blue}\ref{AlphaRi}\color{black}) and we obtain $\alpha_V=0.012\pm0.003$, 
in agreement with previous results for non-buoyant vortex rings \citep{Maxworthy1974,GlezerColes1990}. The largest $\alpha^\prime$ values are reached for the \color{black}largest $Ri$ values and correspond to experiments that have reached a turbulent thermal regime, with $\alpha^\prime$ close to $\alpha_T$ and $Ri$ close to $Ri_T$ ($Ri_T = 0.7\pm 0.2$ in our Immersed experiments). 
At intermediate $Ri$ values ($\sim 0.4\pm0.2$), $\alpha^\prime$ is in the range $0.05-0.17$. 
A best fit of the form $\alpha^\prime = \alpha_V + (\alpha_T - \alpha_V) (Ri/Ri_T)^{\beta}$ for immiscible experiments yields $\beta = 1.2\pm 0.2$, which is compatible with $\beta = 1$ and in agreement with (\ref{AlphaRiP}). 

\color{black}
\subsubsection{\color{black}Descent trajectory}
\color{black}M\color{black}easured distance $\tilde{z}-\tilde{z}_0$ between vortex ring centroid and initial depth for the turbulent thermals (Immersed configuration) is compared with theoretical predictions obtained by numerical integration of equations (\ref{Momentum2})-(\ref{Rz2}) for a constant $\alpha^\prime$ value, as measured in our experiments. 
In each experiment, we choose $\tilde{t}_0$ such that $\tilde{z}_0\approx 1$ in order to ensure that the released fluid is entirely out of the tube at $\tilde{t}_0$. 
The corresponding initial conditions $\tilde{r}_0$ and $\tilde{u}_0$ are then extracted from each experiment.
Squares in figure \ref{Zt}(a) illustrates the descent trajectory 
for a given turbulent thermal in the Immersed configuration. During the last phase, 
$(\tilde{z}-\tilde{z}_0)^2$ grows linearly with time, in agreement with the expected asymptotic behavior given by (\ref{Asymptotic2}). The theoretical evolution fits the data shown in figure \ref{Zt}(a) for $C_D^\prime=0.3\pm0.1$ if $k=0$, and $k = 0.18\pm0.1$ if $C_D^\prime = 0$ (solid curves). The large uncertainties on $C_D^\prime$ and $k$ for a single experiment comes from the uncertainty on $\alpha^\prime$. The drag and added mass coefficients, $C_D^\prime$ and $k$, play a symmetric role in the theoretical solution: an increase in $C_D^\prime$ or $k$ causes a decrease in the slope of $(\tilde{z}-\tilde{z}_0)^2$ in the asymptotic regime (figure \ref{Zt}(a)) as expected from equation (\ref{Asymptotic2})-(\ref{Asymptotic3}). The theoretical solution is also sensitive to $c_1$ as shown in figure \ref{Zt}(a). 

The values of $C_D^\prime$, $k$ and $c_1$ required to fit the descent trajectory 
vary between experiments. In $20\%$ of the $20$ Immersed experiments, the measured curve is located above the theoretical curve computed with ($c_1=1$, $k=0$, $C_D^\prime=0$). As negative values for $k$ or $C_D^\prime$ are not physical, these results require $c_1<1$. In those experiments, $c_1$ ranging from $0.8$ to $0.9$ fits the data, corresponding to an overestimation of the volume \color{black}{up to $25\%$ and an overestimation of the equivalent radius up to $8\%$}. \color{black} In the other Immersed experiments the values of $C_D^\prime$ and $k$ required to fit the observed descent trajectory 
vary from $0$ to about $0.5$. 
Figure \ref{Zt}(b) illustrates the large variability in $C_D^\prime$ and $k$: since the experiments shown have similar $\alpha^\prime$ values, the differences in terminal slope come from differences in $C_D^\prime$ and $k$. The latter coefficients are similar in our miscible fluid experiments and the \color{black}descent trajectory is qualitatively very similar with miscible and immiscible fluids (figure \ref{Zt}(b)). 

%
%
%
 
\begin{figure}
\begin{center}
\includegraphics[width=13.6cm]{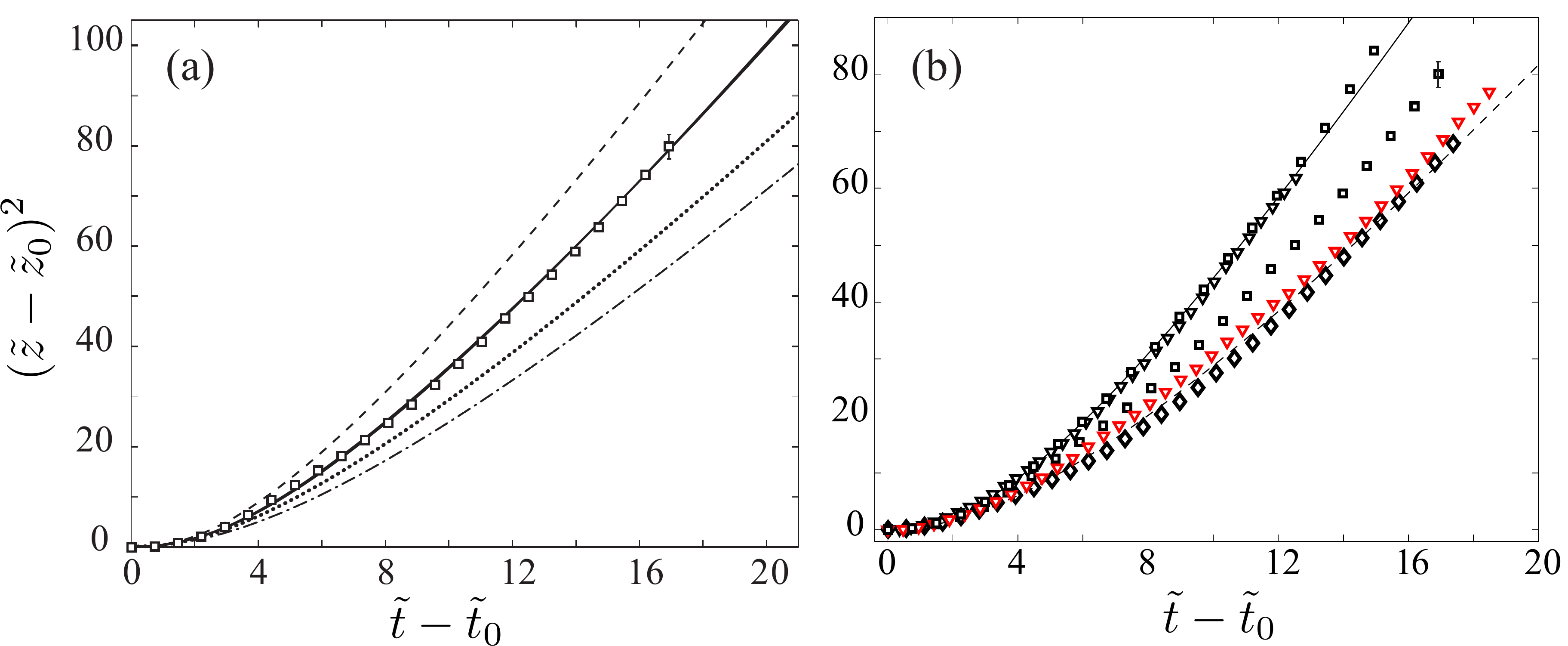}
\end{center}
\caption["Short" caption without tikz code]{
\color{black}Square of dimensionless distance of the vortex ring's centroid from the initial depth as a function of dimensionless time for turbulent thermals in the Immersed configuration. \emptysquare{black}, $\DP \approx 0.92$; \emptydiamond{black}  , $\DP \approx 0.54$;  \emptytriangle{black} , $\DP \approx 0.22$;  \emptytriangle{red} , miscible fluids with $\DP \approx 0.19$. \color{blue}{Characteristic error bars associated with measurement uncertainties are shown.} \color{black} Curves are theoretical solutions obtained by numerical integration of equations (\ref{Momentum2})-(\ref{Rz2}) for constant $\alpha^\prime$ values. (a) Solid curves: two theoretical solutions (indistinguishable from each other) with ($c_1=1$, $C_D^\prime=0$, $k=0.18$) and ($c_1=1$, $C_D^\prime=0.3$, $k=0$), \color{blue}{the mean deviation relative to the experimental curve is less than $0.06\%$}. \color{black} Dotted curve: $c_1=1$, $C_D^\prime=1$, $k=0$. Dot-dashed curve: $c_1=1$, $C_D^\prime=0$, $k=1$. Dashed curve: $C_D^\prime=0$, $k=0.18$, $c_1=0.7$. 
Theoretical curves are computed for $\alpha^\prime = 0.25$, which is the value measured in the experiment shown in (a). (b) 
Immersed experiments in which $\alpha^\prime$ takes similar values in the range $0.23-0.25$. 
Theoretical solutions with ($c_1=1$, $C_D^\prime=0$, $k=0$, $\alpha^\prime=0.24$) and ($c_1=1$, $C_D^\prime=0.35$, $k=0.35$, $\alpha^\prime= 0.24$) are shown by solid and dashed curves respectively. }
\label{Zt}
\end{figure}

In Surface experiments (buoyant vortex rings), our results on the entrainment coefficient (figure \ref{AlphaRi}), combined with theoretical predictions, require $\alpha^\prime$ to vary with time, as already argued in 6.1. Thus, a parameterization such as (\ref{AlphaRiP}) is required for a self-consistent model that predicts the descent trajectory. 
Theoretical solutions obtained by numerical integration of equations (\ref{Momentum2})-(\ref{Rz2}), coupled with parameterization (\ref{AlphaRiP}), fit the $16$ Surface experiments used in this section with $C_D^\prime = 0.6\pm0.3$ and $k=0.4\pm 0.4$, indicating that this model of buoyant vortex ring is consistent with our measurements. $Ri_T$, required in parameterization (\ref{AlphaRiP}), is estimated in each experiment from local measurements of $\alpha^\prime$ and $Ri$. We use the values of $\alpha_T$ and $\alpha_V$ that have been obtained in \S\ref{Coef}, \color{blue}{and we choose the initial time $\tilde{t}_0$ such that no air bubbles remain inside the vortex ring.} 
\color{black}{Figure \ref{Zt_Surf} illustrates the agreement between theoretical and experimental results for a single Surface experiment. 
When using parameterization (\ref{AlphaRiP}), the best-fit theoretical curve is obtained for $C_D^\prime=0.7\pm0.1$ and $k=0.4\pm 0.2$ (with $c_1=1$, $\alpha_V=0.012\pm0.003$ and $\alpha_T=0.23\pm0.06$). $\alpha^\prime$ varies from $0.04$ to $0.14$ in this theoretical solution (figure \ref{Zt_Surf}(b)). The uncertainties on $C_D^\prime$ and $k$ in a single experiment are mainly due to uncertainties on $\alpha_T$, $\alpha^\prime$ and $Ri$. Note that the fit between the data and the theoretical solution is also good with a constant $\alpha^\prime$ value (figure \ref{Zt_Surf}(a)). \color{black}

\color{black}T\color{black}he values we have found for ($C_D^\prime$, $k$, $c_1$), as well as their large variability, are also consistent with results from previous studies. \citet{Ruggaber2000} reports negative values for $C_D^\prime$ and $k$ in turbulent particle clouds, which would be explained by $c_1<1$ in our formalism. The results by \citet{Bush2003} from particle cloud experiments and by \citet{Maxworthy1974} from non-buoyant vortex rings suggest values of $C_D^\prime$ and $k$ small \color{black}compared to $0.5$. 
Translated into our formalism, results of \citet{Gan2012} for non-buoyant vortex rings yield $k\approx 1$ and $C_D^\prime$ of order $0.05$. Although \citet{Thompson2000} do not include the drag coefficient in their model, they report a mean $k$ value of $0.25$ and their data suggest that $k$ ranges from negative values to values close to $0.8$.  

\begin{figure}
\begin{center}
\includegraphics[width=13.6cm]{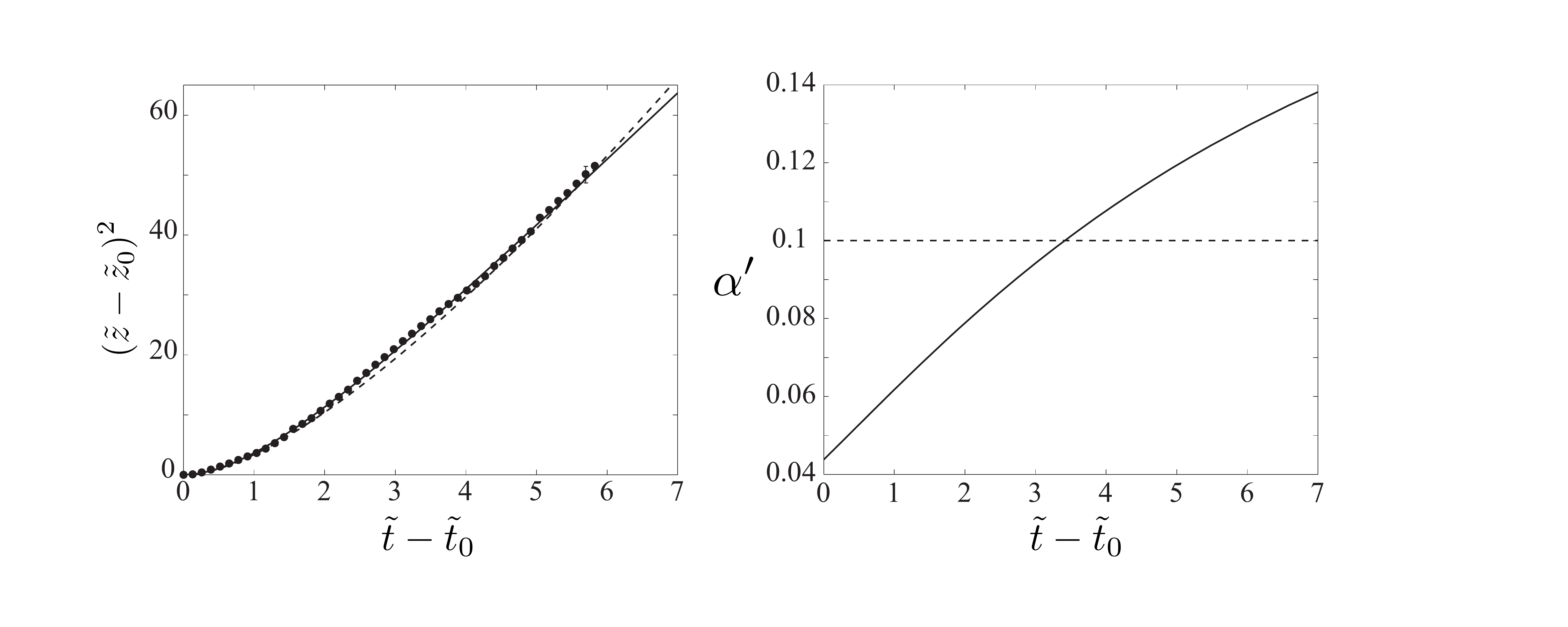}
\end{center}
\caption["Short" caption without tikz code]{
(a) Square of dimensionless distance of the vortex ring's centroid from the initial depth as a function of dimensionless time.  \colourcircle{black}, buoyant vortex ring in the Surface configuration with $\DP \approx 0.03$. \color{blue}{Characteristic error bars associated with measurement uncertainties are shown.} \color{black} Black curves are theoretical solutions obtained by numerical integration of equations (\ref{Momentum2})-(\ref{Rz2}) with either a constant $\alpha^\prime$ value (dashed curve) or $\alpha^\prime$ that varies with time according to (\ref{AlphaRiP}) (solid curve). $\tilde{t}_0$ is chosen such that $\tilde{z}_0\approx 2$. Dashed curve: $\alpha^\prime=0.1$, $C_D^\prime=k=0.34$, $c_1=1$, \color{blue}{with a mean deviation relative to the experimental curve equal to $0.06\%$}. \color{black} Solid curve: $k = 0.45$, $C_D=0.68$, $c_1=1$, \color{blue}{with a mean deviation relative to the experimental curve equal to $0.02\%$}. \color{black} (b) Entrainment coefficient as a function of dimensionless time in the two theoretical solutions shown in (a). }
\label{Zt_Surf}
\end{figure}

\section{Discussion}
\subsection{Discussion of experimental results}

We find that, in agreement with the literature on fluid fragmentation, and especially on drop breakup at low $Oh$ \citep{Hinze1955,PilchandErdman,Gelfand1996,Guildenbecher2009}, the Weber number is the control parameter governing regime transitions in \color{black}{our experiments, whereas $\DP$ has an influence mostly within the fragmentation regime (figure \ref{DP_all}). 


The vortex ring destabilization regime found in this study is morphologically different from the regime observed by \cite{Baumann1992}. 
In their study, immiscible vortex rings are rather viscous ($Re\leq61$) whereas, in the present study, $Re\geq 10^3$ in most immiscible vortex rings, closer to inviscid dynamics. 
The destabilization of vortex rings in \citet{Baumann1992} is interpreted as a manifestation of RTI and is morphologically similar to the instability observed in miscible fluids when a drop of a heavier liquid falls \color{black}inside a lighter one \citep{Kojima1984,Arecchi1989,Arecchi1991,Bua2005}. 
The centrifugal to gravitational acceleration ratio is much smaller than $1$ in the vortex rings of \citet{Baumann1992}, indicating that RTI are mainly driven by gravity. The same ratio (roughly estimated from video images) reaches values of about $0.5$ in some of our immiscible vortex rings, demonstrating that the destabilizing mechanisms can not be identical to those in \citet{Baumann1992}. 


A progressive transition leads to a turbulent regime that is observed for \hbox{$\Web\gtrsim100-200$} in both Immersed and Surface experiments. We emphasize that in $(\Bo, \DP)$ space the transition to turbulence would occur at different parameter values in Immersed and Surface experiments (surface experiments at $P\approx 0.03$ and $\Web \geq 100$ have a rather low $\Bo$ value compared to other turbulent experiments). 

In our turbulent experiments, the turbulent entrainment concept describes the large-scale evolution of the released fluid even before breakup occurs, for distances smaller than $4.5-7.5$ initial radii. 
At this stage, both the ambient and released fluids form continuous, non-dispersed phases. 

It is not clear whether our turbulent regime corresponds to the regime described by \cite{YangYang}. 
In our experiments, the entrainment coefficient 
decreases when reducing the local Richardson number $Ri$. 
\cite{YangYang} report that the entrainment coefficient grows as the square root of the Weber number, at similar $Bo$ values. Noting that $Ri=O(Bo/We)$, their results seem to be at variance with ours, and may indicate that the fully turbulent regime has not been reached in their experiments.

\color{blue}{
In our positively buoyant vortex rings, the amplitude of baroclinically-generated vorticity, correlated with $Ri$, affects 
the value of the entrainment coefficient and the amplitude of the detrainment process by which vortical structures are shed into the wake. These results are in agreement with previous studies that show that the amplitude and sign of baroclinically-generated vorticity generally control the dynamics in buoyant shear flows. \citet{MaruganCruz2009,MaruganCruz2013} have shown that the dimensionless pinch-off time of the leading vortex ring varies with the Richardson number in a starting negatively buoyant jet. Comparing the scaling law for pinch-off time in non-buoyant starting jets \citep{Gharib1998} and in positively buoyant starting plumes \citep{Shusser2000,Pottebaum2004}, it is found that the pinch-off time should vary with the Richardson number in positively buoyant starting jets. 
The penetration depth or the entrainment rate in negatively \citep[e.g.][]{Turner1966,Carazzo2010} and positively \citep{Fisher1979,Kaminski2005} buoyant jets, and at a stratified interface \citep{Linden1973,Baines1975,Cotel1997,CotelBreidenthal1997,Friedman1999,Friedman2000,ZhangCotel2000} also depends on a Richardson number.}

\color{black}

For sufficiently large $We$, the dimensionless breakup length remains in the range $4.5-7.5$, with no significant variations when increasing further $We$. These results suggest either that $L_B$ tends towards a constant in the limit of large $We$, or that it exhibits a weak dependence on $We$. 
Further investigations of the turbulent regime at $We\geq 10^3$ are required to test these hypotheses. 
However, a result shared by all these studies is that, for large enough $We$ 
and $Oh\ll1$, the dimensionless breakup time shows no 
significant dependence on $We$, which is in agreement with our data.    


\subsection{Geophysical implications}

\subsubsection{Earth's core formation}
%





The migration and fragmentation of liquid metal in fully liquid silicate magma oceans 
is likely to have played an important role in determining the final composition of Earth's core and Earth's mantle since the small-scale intermingling between metal and silicates allowed for chemical equilibration \citep{DahlStevenson2010}. 
After an impact between differentiated (i.e. formed of a silicate mantle and metallic core) planetary embryos, the initial radius and post-impact velocity of released metal blobs are expected to be in the range $50-500$ km and $0.1-10$ km$\cdot$s$^{-1}$, respectively \citep{Rubie2003,Canup2004,Deguen2011}. 
The depth of the magma ocean was, at most, of the same order of magnitude as the depth of the present Earth's mantle, i.e. about $3000$ km. 
Thus, the characteristic time scale for the first stages of metal migration in a magma ocean did not exceed a few hours, suggesting that the effects of rotation can be neglected at first order. 
The density of liquid metal and liquid silicates at magma ocean depths are typically in the range \color{blue}{$\mathbf{7000-9000}$ kg$\cdot$m$^{-3}$ \citep{Morard2013}} \color{black} and $3000-4000$ kg$\cdot$m$^{-3}$ \citep{Miller1991}, respectively. 
The interfacial tension between liquid metal and liquid silicates is expected to be of order $1$ J$\cdot$m$^{-2}$ \citep{ChungCramb2000}, although it varies significantly with temperature, light element content and pressure \citep{Terasaki2012}. The viscosity of a fully liquid magma ocean is at most of order \hbox{$0.1$ Pa$\cdot$s} \citep{Liebske2005,KarkiStixrude2010} while it is likely to be in the range \hbox{$10^{-3}-5\times 10^{-2}$ Pa$\cdot$s} for liquid metal \citep{Funakoshi2010,DeWijs1998}. With the above estimates, we expect $We\gtrsim 10^{12}$, $Oh\lesssim 10^{-5}$, $Bo\gtrsim10^{13}$ and $Re\gtrsim10^{11}$ following an impact (using the equivalent radius of the metal blob as a length scale), 
with a \color{blue}{normalised density difference }\color{black}$\DP$ of order $1$ for the metal-silicate system. 

Although our laboratory experiments \color{blue}assume uniform background conditions and \color{black} are far from reproducing post-impact conditions that prevailed during planetary formation, they give insights into the flow regime associated with the fragmentation of metal blobs in a fully liquid magma ocean.
If we locate proto-planets, including proto-Earth, in the regime diagram of figure \ref{DP_all}, they would be close to the line $P=1$ at $We\geq 10^{12}$, indicating that the geophysical flows of interest are located well above the onset of the turbulent regime at $\Web \sim 200$. 
Thus, even if the largest $\Web$ values reached in our experiments are more than $9$ orders of magnitude smaller than in the geophysical system, 
we have explored the regime that is relevant 
for core formation, \color{blue}and in which the large-scale flow can be described by considering the limit of zero surface tension. 
\color{black} 

Morphologically, the turbulent fragmentation regime is very different from the classic picture found in the literature on planet formation, where a cascade of fragmentation events progressively leads to smaller and smaller fragments \citep{Rubie2003,Samuel2012}, eventually resulting in an \textit{iron-rain} falling in a magma ocean \citep{Ichikawa2010}. It is also different from erosion models \citep{DahlStevenson2010} where metal-silicate intermingling occurs only at the metal blob boundary. Our experiments rather suggest that metal fragmentation occurs in a turbulent immiscible vortex ring which grows by entrainment of silicates and where metal and silicates are intimately intermingled in the whole ring volume. 
Quantitative implications of those findings for mantle and core geochemistry are further discussed in a companion paper \citep{Deguen2013} where a model of chemical equilibration between metal and silicates in a magma ocean is developed. \color{blue}In particular, it is shown that surface tension controls the small-scale flow and is the essential limiting factor for chemical transfers. \color{black}





The integral model proposed in \S6 is expected to apply for the migration of a metal blob in a fully liquid magma ocean. It provides the 
descent trajectory of the metal-silicate mixture and the amount of silicates that are mixed with metal. The latter can be deduced from equation (\ref{Rz}) and depends on the value of the entrainment coefficient $\alpha^\prime$. For distances much larger than the Morton length $L_M$, the entrainment coefficient is equal to its value in turbulent thermals, i.e. $\alpha_T = 0.23\pm0.06$. For distances of the same order of magnitude as $L_M$ or smaller, the value of $\alpha^\prime$ depends on the local Richardson number and it takes values between $\alpha_V = O(0.01)$ and $\alpha_T$. 
With $g=5$ m$\cdot$s$^{-2}$, the initial Richardson number for a $100$ km sized metal blob can reach values 
in the range $10^{-3}-10$ and in the range $10^{-2}-100$ 
for a $1000$ km sized blob. For initial Richardson numbers equal to $1$ or larger, no significant departure from $\alpha^\prime = \alpha_T$ can be caused by $Ri$ variations. 
For cases where the initial Richardson number is of order $10^{-3}-10^{-2}$, $\alpha^\prime$ is expected to be initially close to its value in non-buoyant vortex rings, $\alpha_V =O(0.01)$. In such cases, $L_M$ is of about $100$ initial radii, which is always larger than the magma ocean depth, suggesting that the entrainment coefficient is influenced by $Ri$ during the entire fall. 
Thus, a large post-impact velocity can decrease the rate of entrainment by a factor $10$, reducing the total volume of silicates mixed with metal during its fall by a factor $10^3$. This effect should be taken into account in models of metal-silicate equilibration.

\color{blue}{
Another consequence of low $Ri$ values is the detrainment of vortical fluid ejected into a wake, as qualitatively supported by our experiments. This process should be taken into account in models of metal-silicate equilibration following impacts where the mass of metal ejected into the wake of the sinking core was significant. 
However, measurements of detrainment in buoyant vortex rings at lower $Ri$ are required to draw any firm conclusions.\\
}
\color{black}

%
%
As discussed in the previous section, it is possible that the dimensionless breakup length remains constant when $We$ increases, taking values in the range $4.5-7.5$ initial radii. Then, breakup would occur during the fall for blobs with an equivalent radius at least $10$ times smaller than the magma ocean depth. In the case of giant impacts, the size of the impactor core is of the same order of magnitude as the depth of the magma ocean ($O(1000)$km) and it is possible that 
breakup does not begin before the liquid metal reaches the bottom of the magma ocean. A secondary impact at the bottom of the magma ocean, with either liquid metal (if the magma ocean depth is equal to the mantle depth) or solid silicates, would then play a major role in the fragmentation process.    

\color{blue}{
\subsubsection{Deepwater Horizon disaster}

The blowout at the Macondo well on the floor of the Gulf of Mexico that followed the April 20, 2010 sinking of the Deepwater Horizon (DH) platform resulted in the largest offshore oil spill in history \citep{McNutt2012}. Observations of the fragmentation of petroleum during the DH oil spill and the subsequent migration of the oil drop clouds so produced offer a basis for predicting the fate of future hydrocarbon spills in the deep marine environment. 

Initially, the DH oil emission came from two main leaks on the ocean floor separated horizontally by about $250$ m at a depth of about $1.5$ km, and forming two multiphase plumes \citep{Socolofsky2011,Camilli2012}. By virtue of the stratification of the water column above the spill, some of the hydrocarbons were trapped between $1000$ and $1300$ m depth \citep{Reddy2012,Camilli2010,Joye2011}. In general, in a stratified environment, the density of a rising plume increases due to entrainment of ambient fluid, and because the density of the ambient fluid decreases with height, the plume eventually stops at its level of neutral buoyancy, as defined by \citet{Morton1956}. 
In the case of the DH plumes, the dynamics were further complicated by nonlinear stratification in the water column and by the presence of multiple phases, including oil and various aqueous and gas phases \citep{Camilli2010,Socolofsky2011,Adalsteinsson2011}. 

The DH disaster has fundamental differences with our experiments, including the configuration (maintained source rather than instantaneous release), the presence of stratification (absent in our experiments), the presence of gases, and the partial aqueous solubility of the released fluids \citep{Reddy2012}. Thus, any quantitative comparison would require additional experiments with imposed input flow rate, while including the above additional ingredients. Despite these differences, however, our results give insights into the relevant regime for the fragmentation and early evolution of the spilt oil. Given the physical properties of the oil at its release point \citep{Socolofsky2011,Camilli2012,RapportDH}, a mean section area for each plume in the range $0.99-1.10$ m$^2$ and a characteristic velocity at the source of order $0.1$ m$\cdot$s$^{-1}$ \citep{Camilli2012}, we estimate that $We$ and $Oh$ are of order $400$ and $10^{-3}$, respectively, for the oil-water system. Our results imply that the oil fragmentation regime is therefore fully turbulent (figure \ref{DP_all}), suggesting that entrainment at the oil-water interface can be described according to the same turbulent entrainment hypothesis used in our study.

} \color{black}

\section{Conclusion}

We have described a series of experiments on liquid-liquid fragmentation at low $Oh$, varying the \color{blue}{normalised density difference }\color{black} ($ 0.03\leq \DP\leq 0.95$) and the Weber number ($ 1 \lesssim \Web \lesssim  10^3$). We have shown that the typical stages of any fluid fragmentation process are found in our experiments: from the deformation and destabilization of the released fluid to the formation of liquid filamentary structures that break by capillary instabilities and form fragments. We have studied the destabilization and macroscopic evolution of the released fluid, from which fragmentation regimes were characterized. 

We have found that, at low and intermediate Weber numbers, the fragmentation regime is very sensitive to the 
release conditions (Immersed versus Surface) and a wide variety of regimes is identified. Most of those fragmentation regimes are influenced by early deformations, which result from a competition between growth of RTI and roll-up of a vortex ring.

At high Weber numbers ($\Web \gtrsim 200$) a turbulent flow regime is reached and the large-scale flow shares common features in all the experiments: the released fluid is contained inside a coherent structure whose shape is, at first order, self-similar during the fall and which grows by turbulent entrainment of ambient fluid. To our knowledge, we have reported the first visualizations of immiscible turbulent thermals and immiscible turbulent vortex rings in a non-dispersed medium. Previously published models based on the turbulent entrainment concept have been extended to the general case of buoyant vortex rings. Our results indicate a positive correlation between the entrainment coefficient and the local Richardson number. 
The consistency between experimental and theoretical results, and between results from miscible and immiscible fluid experiments, supports that the turbulent entrainment concept can be applied in the context of non-dispersed immiscible fluids at large Weber and Reynolds numbers.


Important information such as drop size distributions and small-scale mechanisms leading to breakup in immiscible turbulent thermals and turbulent vortex rings require further investigation, along with 
additional experimental studies to confirm the precise relationship between the entrainment coefficient and the local Richardson number in miscible and immiscible systems. \\

%

%
%
%
%
%
%
The experiments in this study were supported by U.S. National Science Foundation Grant EAR-110371 to Johns Hopkins University.

\appendix
\section{Preliminary processing}
\label{PreliminaryProc}
Binary images are obtained by subtracting the back field image, taken before the release of dense fluid, to each video frame.  
Then we select an appropriate pixel intensity threshold $I_c$, above which the pixel intensity is set to $1$, and $0$ otherwise. 
The threshold is chosen as $I_c = c\cdot I_{noise}$ where $c$ is a constant specified by the operator and $I_{noise}$ the standard deviation to $0$ of the back field noise. If $X$ is the set of pixels with negative intensity values after subtraction of the back field, $I_{noise}$ is given by
\begin{eqnarray}
I_{noise} = \sqrt{\dfrac{1}{N_{X}}\sum_{(i,j)\in X}I_{i,j}^2},
\label{SigmaNoise}
\end{eqnarray}
where $N_X$ is the number of pixels in $X$. We use \color{blue}the color channel \color{black} in which the absorption of light by the released fluid is the largest (i.e. red \color{blue}color channel \color{black} for blue-dyed fluid). The value of $c$ is chosen such that the output variables that are eventually obtained from binary images ($z$, $u$, $r$, $L_B$, $\alpha^\prime$ defined in \S2, \S2, \S2, \S5, \S6, respectively) do not vary significantly with $c$. Sensitivity of output variables to $c$ are included in measurement uncertainties. $c$ is held constant for a particular group of experiments (same lighting conditions and same fluids). 
%

\section{Turbulent entrainment model: closed-form solutions}
\label{TheoreticalSol}

In the Boussinesq limit $P\rightarrow 0$, the solution (\ref{GeneralSolution}) to equation (\ref{Momentum3}) takes the following closed-form expression:

\begin{eqnarray}
\tilde{u}^2  = \dfrac{\tilde{r}_0^{3\gamma}}{\tilde{r}^{3\gamma}}\left\{ \tilde{u}_0^2 - \dfrac{1}{2\alpha^\prime}\left \lbrack \dfrac{3C_D^\prime}{16\alpha^\prime} +(1+k)c_1 \right \rbrack^{-1} \left \lbrack \dfrac{1}{\tilde{r}_0^2} - \dfrac{\tilde{r}^{3\gamma-2}}{\tilde{r}_0^{3\gamma}}\right \rbrack  \right \}
\label{Appendix_SolBouss}
\end{eqnarray}
where $\tilde{r}=\tilde{r_0}+\alpha^\prime(\tilde{z}-\tilde{z}_0)$. 
%
%
%
%
%

A closed-form solution also exists for $C_D^\prime=0$ and is given by

\begin{eqnarray}
\tilde{u}^2  = \left \lbrack \dfrac{P+(1+k)c_1\tilde{r}_0^3}{P+(1+k)c_1\tilde{r}^3}\right \rbrack ^2 \left \lbrack \tilde{u}_0^2+  \dfrac{2P(\tilde{r}-\tilde{r}_0)}{\alpha^\prime(P+(1+k)c_1\tilde{r}_0^3)^2}
+  \dfrac{c_1(1+k)(\tilde{r}^4-\tilde{r}_0^4)}{2\alpha^\prime(P+(1+k)c_1\tilde{r}_0^3)^2} \right \rbrack
\label{Appendix_SolCd}
\end{eqnarray}
where $\tilde{r}=\tilde{r_0}+\alpha^\prime(\tilde{z}-\tilde{z}_0)$. The first term within the second brackets in equation (\ref{Appendix_SolCd}) is due to the initial momentum of the vortex ring, the second term to departures from the Boussinesq approximation and the third term is related to buoyancy forces.

\bibliographystyle{jfm}
\bibliography{Bib_2}

\end{document}